\documentclass[twocolumn]{aastex6}
\usepackage{times}
\usepackage{amsmath}
\usepackage{textcomp}
\usepackage{amssymb}
\usepackage{color}

\newcommand{\msun}{\ensuremath{M_{\odot}}}
\newcommand{\lum}{erg\,s$^{-1}$}
\newcommand{\fermi}{{\it Fermi}}
\newcommand{\nustar}{{\it NuSTAR}}

\newcommand{\swift}{{\it Swift}}
\newcommand{\phflux}{\mbox{${\rm \, ph \,\, cm^{-2} \, s^{-1}}$}}
\newcommand{\ergflux}{\mbox{${\rm \, erg \,\, cm^{-2} \, s^{-1}}$}}
\newcommand{\gm}{$\gamma$}

\slugcomment{submitted to ApJ}

\shorttitle{DA 193}
\shortauthors{Paliya et al.}

\begin{document}

\title{Detection of A Gamma-ray flare from the high-redshift blazar DA 193}

\author{Vaidehi S. Paliya$^{1,2}$, M. Ajello$^2$, R. Ojha$^3$, R. Angioni$^{4,5}$, C. C. Cheung$^6$, K. Tanada$^{7}$, T. Pursimo$^{8}$, P. Galindo$^{8}$, I. R. Losada$^{8}$, L. Siltala$^{8}$, A. A. Djupvik$^{8}$, L. Marcotulli$^{2}$, D. Hartmann$^{2}$} 
\affil{$^1$Deutsches Elektronen Synchrotron DESY, Platanenallee 6, 15738 Zeuthen, Germany}
\affil{$^2$Department of Physics and Astronomy, Clemson University, Kinard Lab of Physics, Clemson, SC 29634-0978, USA}
\affil{$^3$NASA Goddard Space Flight Center, Greenbelt, MD 20771, USA}
\affil{$^4$Max-Planck-Institut f\"ur Radioastronomie, Auf dem H\"ugel 69, D-53121 Bonn, Germany}
\affil{$^5$Institut f\"ur Theoretische Physik und Astrophysik, Universit\"at W\"urzburg, Emil-Fischer-Str. 31, 97074 W\"urzburg, Germany}
\affil{$^6$Space Science Division, Naval Research Laboratory, Washington, DC 20375-5352, USA}
\affil{$^7$Research Institute for Science and Engineering, Waseda University, 3-4-1, Okubo, Shinjuku, Tokyo, 169-8555, Japan}
\affil{$^8$Nordic Optical Telescope, Nordic Optical Telescope Apartado 474E-38700 Santa Cruz de La Palma, Santa Cruz de Tenerife, Spain}
\email{vaidehi.s.paliya@gmail.com}

\begin{abstract}
High-redshift ($z>2$) blazars are the most powerful members of the blazar family. Yet, only a handful of them have both X-ray and \gm-ray detection, thereby making it difficult to characterize the energetics of the most luminous jets. Here, we report, for the first time, the \fermi-Large Area Telescope detection of the significant \gm-ray emission from the high-redshift blazar DA 193 ($z=2.363$). Its time-averaged \gm-ray spectrum is soft (\gm-ray photon index = $2.9\pm0.1$) and together with a relatively flat hard X-ray spectrum (14$-$195 keV photon index = $1.5\pm0.4$), DA 193 presents a case to study a typical high-redshift blazar with inverse Compton peak being located at MeV energies. An intense GeV flare was observed from this object in the first week of 2018 January, a phenomenon rarely observed from high-redshift sources. What makes this event a rare one is the observation of an extremely hard \gm-ray spectrum (photon index = $1.7\pm0.2$), which is somewhat unexpected since high-redshift blazars typically exhibit a steep falling spectrum at GeV energies. The results of our multi-frequency campaign, including both space- (\fermi, \nustar, and \swift) and ground-based (Steward and Nordic Optical Telescope) observatories, are presented and this peculiar \gm-ray flare is studied within the framework of a single-zone leptonic emission scenario. 
\end{abstract}


\section{Introduction}{\label{sec:Intro}}
The extragalactic high-energy \gm-ray sky, as observed by the Large Area Telescope \citep[LAT,][]{2009ApJ...697.1071A} onboard the \fermi~Gamma-ray Space Telescope, is dominated by blazars. These are a peculiar class of active galactic nuclei (AGN) which host powerful relativistic jets aligned close to the line of sight to the observer. Based on the rest-frame equivalent width (EW) of the emission lines, blazars are sub-classified as flat spectrum radio quasars (FSRQs, EW$>$5\AA) and BL Lac objects \citep[][]{1991ApJ...374..431S}. The optical spectrum of a FSRQ exhibits broad or strong emission lines, thus indicating the presence of an efficient accretion process surrounding the central black hole. On the other hand, observations of very weak or no emission lines (EW$<$5\AA) in the optical spectra of BL Lac objects suggests either the dominance of the non-thermal jet emission, i.e. the power-law continuum, over emission lines and/or the inefficient and low accretion and thus a weak broad line region \citep[BLR;][]{2005MmSAI..76...61W}. Indeed, a division between FSRQ and BL Lac objects based on the luminosity of the BLR measured in Eddington units has been proposed by \citet[][]{2011MNRAS.414.2674G} with FSRQs exhibiting $L_{\rm BLR}/L_{\rm Edd}>0.0005$.

The spectral energy distribution (SED) of a blazar is characterized by a double peak structure and is governed by the bolometric luminosity of the source, also known as the blazar sequence (\citealt[][]{1998MNRAS.299..433F,1998MNRAS.301..451G,2001A&A...375..739D}; but see \citealt{2012MNRAS.420.2899G}). It has been observed that the low energy synchrotron peak of luminous blazars (generally FSRQs) lies at sub-mm to infrared (IR) frequencies, whereas, the high energy inverse Compton (IC) emission peaks at $\sim$MeV energies. In general, high-redshift ($z>2$) blazars are the most luminous ones. Therefore, in the broadband SED of high-redshift blazars, thermal emission from the accretion disk is visible in the optical-UV band due to shift of the synchrotron peak to lower frequencies \citep[e.g.,][]{2010MNRAS.405..387G}. Moreover, due to their IC emission peak lying at MeV energies, these distant sources are also called as `MeV' blazars \citep[][]{1995A&A...293L...1B}. 

Lacking an MeV all-sky instrument, the most efficient domain where to detect and study high-redshift blazars is the hard X-ray ($>$10 keV) band. In this energy range, these objects display remarkably hard spectra (power-law photon indices $\lesssim$1.6) that easily distinguish them from the other, more common, AGN. {\it Swift}-Burst Alert Telescope \citep[BAT;][]{2005SSRv..120..143B} has detected 22 such objects at $z\geq$2, all with a luminosity Log$L_{\rm X}\geq$47.5\,erg s$^{-1}$ \citep[][]{2018ApJS..235....4O}. Many of them were found to have larger-than-average jet powers ($>10^{47}$ \lum), accretion luminosities, Compton dominance (which is the ratio of IC to synchrotron peak luminosities), and black hole masses \citep[][]{2010MNRAS.405..387G,2015ApJ...807..167T,2016MNRAS.462.1542S}.

Though hard X-ray observing facilities, e.g. \nustar~\citep[3$-$79 keV;][]{2013ApJ...770..103H}, are better suited to study high-redshift blazars, \fermi-LAT observations are still crucial to constrain the falling part of the IC spectrum and thus the location of the IC peak. More importantly, an unambiguous confirmation of the blazar nature of a high-redshift radio-loud quasar itself can be made by the \fermi-LAT detection which provides strong evidences in support of the closely aligned relativistic jet. In this regard, recently released Pass 8 photon data set, with an improved event-level analysis \citep[][]{2013arXiv1303.3514A}, substantially enhances the LAT capability to detect spectrally soft, potentially high-redshift blazars that host massive black holes and the most powerful relativistic jets. Since blazars are highly beamed, the detection of a single blazar implies the existence of hundreds of quasars with similar properties, and at the same redshift, but with jets pointed in other directions. Therefore, the discovery of each new high-redshift blazar provides crucial constraints on the space density of massive black holes, hosted in radio-loud systems, in the early Universe \citep[][]{2009ApJ...699..603A,2010MNRAS.405..387G,2017ApJ...837L...5A}.

We are studying high-redshift blazars with the motivation to characterize their physical properties using \fermi-LAT and other multi-frequency observations \citep[see,][]{2015ApJ...804...74P,2016ApJ...825...74P,2016ApJ...826...76A,2017ApJ...834...41K,2017ApJ...837L...5A,2017ApJ...839...96M,2017ApJ...851...33P,2018ApJ...859...80K}. In this work, we present the results of our study on another high-redshift blazar DA 193 \citep[also known as 0552+398; $z=2.363$,][]{1976ApJS...31..143W,1999ApJ...514...40M} which we have found as a new \gm-ray emitting object through our detailed \fermi-LAT analysis. Interestingly, this source exhibited a GeV flare in 2018 January \citep[][]{2018ATel11137....1A}. To investigate and study this peculiar event in detail, we triggered target of opportunity (ToO) observations from the {\it Neil Gehrels \swift~observatory} and \nustar, and also optical polarimetric and photometric measurements from the Steward observatory and Nordic Optical Telescope (NOT), respectively. In particular, optical followups, including polarization measurements, are helpful in determining the relative dominance of the synchrotron emission and the accretion disk radiation used for the SED modeling. In Section \ref{sec2}, basic information of DA 193 are briefly described. The steps of data reduction and analysis are elaborated in Section \ref{sec3}. Results are presented and discussed in Section \ref{sec4} and we summarize them in Section \ref{sec5}. Throughout, we assume a flat cosmology with $H_0=67.8$ km s$^{-1}$ Mpc$^{-1}$ and $\Omega_{\rm M}=0.308$ \citep[][]{2016A&A...594A..13P}.
 
\section{Basic Multiwavelength Information}{\label{sec2}} 
DA 193 lies close to the Galactic anti-center (Galactic longitude and latitude: 171$^{\circ}$.647 and 7$^{\circ}$.284, respectively) and is a bright radio quasar \citep[$F_{\rm 1.4~GHz}=$1516 mJy;][]{1998AJ....115.1693C}. It was the subject of various radio studies due to its peculiar Gigahertz peak spectrum \citep[e.g.,][]{1981A&A....96..365S,1983ApJ...271...44S,1990A&AS...84..549O}. It is one of the MOJAVE monitored blazars and the detection of superluminal motion has been reported at 15 GHz \citep[apparent jet velocity = 1.592$\pm$0.096 c,][]{2016AJ....152...12L}. Interestingly, by studying multi-epoch Very Long Baseline Interferometry observations, \citet[][]{2001A&A...380..123W} proposed the detection of an accelerated jet.

DA 193 is well detected in 2MASS \citep[][]{2006AJ....131.1163S} and {\it WISE} \citep[][]{2010AJ....140.1868W} surveys. The optical spectroscopic observations of this object led to the detection of broad emission lines which supports the case for an extremely massive black hole at the center \citep[$M_{\rm BH}\approx 5\times10^9$ \msun;][]{2003ApJ...593L..11Y,2012RMxAA..48....9T}. Moreover, optical polarimetric measurements from McDonald observatory resulted in the detection of low polarization of $\sim$1.5\% \citep[][]{2011ApJS..194...19W}.

DA 193 has also been studied at X-ray frequencies to understand the soft X-ray absorption typically observed in many high-redshift radio-loud quasars \citep[e.g.,][]{2013ApJ...774...29E,2018A&A...616A.170A}. This object is luminous at hard X-rays and exhibits a relatively flat X-ray spectrum above 10 keV \citep[$L_{\rm 14-195~keV}=10^{48}$ \lum, $\Gamma_{\rm 14-195~keV}=1.5\pm0.4$;][]{2018ApJS..235....4O}, typical of a powerful MeV blazar. However, it is not present in any of the published \gm-ray catalogs.
 
\section{Data Reduction}{\label{sec3}}
\subsection{Gamma-ray Analysis}
We analyze \fermi-LAT Pass 8 data of DA 193 following the standard data reduction procedure\footnote{https://fermi.gsfc.nasa.gov/ssc/data/analysis/documentation/} and briefly describe it here. In the energy range of 0.1-300 GeV, we consider only SOURCE class events ({\tt evclass=128}) covering the period MJD 54683$-$58137 (2008 August 5-2018 January 19) and adopt a relational filter {\tt `DATA\_QUAL$>$0 and LAT\_CONFIG==1'} to select good time intervals. Only events passing the cut `zmax$<$90$^{\circ}$' are considered, to avoid Earth limb \gm-ray contamination. A circular region of interest (ROI) is defined with the radius of 15$^{\circ}$ centered at DA 193 \citep[R.A.: 05$^h$55$^m$ 30.806$^s$, J2000; Decl.: +39$^{d}$48$^{\prime}$49.165$^{\prime\prime}$, J2000;][]{2016AJ....152...12L} and a background model is generated consisting of all 3FGL sources \citep[][]{2015ApJS..218...23A} present within the ROI and the Galactic diffuse and extragalactic isotropic background emission templates\footnote{https://fermi.gsfc.nasa.gov/ssc/data/access/lat/BackgroundModels.html} \citep[][]{2016ApJS..223...26A}. The spectral shapes of all the sources are the same as those adopted in the 3FGL catalog and associated parameters are left free to vary when performing the binned likelihood fitting. We determine the significance of the \gm-ray signal by deriving the maximum likelihood test statistic TS=2$\Delta \log (\mathcal{L})$ where $\mathcal{L}$ is the likelihood value between models with and without a \gm-ray source at the position of DA 193. Since the target object is not present in the 3FGL catalog, we parametrize its spectral shape by a power-law model and the prefactor and photon index of the model are allowed to vary during the optimization. 

We follow an iterative approach to search for unmodeled \gm-ray sources present in the data but not in the model, since the time period covered is much longer than that considered in the 3FGL catalog. This is done by generating a residual TS map and scanning it to identify excess emission peaks with TS$>$25 \citep[$\sim$4.2$\sigma$ significance;][]{1996ApJ...461..396M}. Once found, they are modeled with a simple power-law and included in the sky model. This process is repeated until there is no source with TS$>$25 left to model\footnote{We have verified our results by considering the sources not present in 3FGL but in recently released \fermi-LAT 8-year preliminary list (FL8Y) to populate the model file. No significant changes in the optimized spectral parameters are found with respect to that derived from the 3FGL + TS map approach.}. When using the whole data set, a new source is found at the position of DA 193, as fully described in section \ref{subsec:gm}.

To generate the \gm-ray light curves and spectra, we allow the spectral parameters of the source of interest and all other background objects, lying within 10$^{\circ}$ from the ROI center, to vary freely. We compute 2$\sigma$ flux upper limits in time/energy bins in which the source TS is found to be $<$9 or $\bigtriangleup F_{\gamma}/F_{\gamma} > 0.5$, where $\bigtriangleup F_{\gamma}$ is the uncertainty in the 0.1$-$300 GeV \gm-ray flux, $F{\gamma}$. Unless otherwise noted, all the statistical uncertainties are estimated at 1$\sigma$ confidence level.

\subsection{Hard X-ray Analysis}
DA 193 is included in the 105 month \swift-BAT catalog \citep[][]{2018ApJS..235....4O} and we use the publicly available 14$-$195 keV spectrum for SED modeling.

DA 193 underwent a huge GeV flare in the first week of 2018 and a ToO observation with \nustar~was performed on MJD 58126-58127 (2018 January 8-9, PI: V. S. Paliya) to follow the flaring blazar. The net exposure of the observation was $\sim$28 ksec. \nustar~data are reduced following the standard guidelines\footnote{https://heasarc.gsfc.nasa.gov/docs/nustar/analysis/nustar\_swguide.pdf}. In particular, we use the \nustar~Data Analysis Software (NUSTARDAS version 1.8.0) to analyze the Focal Plane Module A (FPMA) and Focal Plane Module B (FPMB) data. Considering the latest \nustar~calibration files (CALDB version 20180419) and standard filtering criteria, the event data files are cleaned and calibrated using the task {\tt nupipeline}. We consider the source region as a circle of 30$^{\prime\prime}$ centered at DA 193 and also use a background (source-free) region of 70$^{\prime\prime}$ from the same chip. We then use the task {\tt nuproducts} to generate source and background spectra and ancillary and response matrix files. The source spectrum is binned to have at least 20 counts per bin using the task {\tt grppha}.

\subsection{Soft X-ray Analysis}
Prior to our three ToO observations in 2018 January, DA 193 was observed by \swift~twice in 2007 and 2010 and thrice in 2011. All of the X-ray Telescope \citep[XRT;][]{2005SSRv..120..165B} observations were taken in the standard photon counting mode (standard grade selection 0$-$12). We downloaded all the available data sets from HEASARC archive and analyze it using the HEASoft (v 6.22) and calibration files updated on 2018 March 5. The event files are cleaned and calibrated using the task {\tt xrtpipeline} and then summed using {\tt xselect}.  We combine exposure maps using the tool {\tt ximage}. To extract the source and background spectra from the summed event file, we consider a circular region of 55$^{\prime\prime}$ radius centered at the target quasar. The background region is taken as annular ring centered at DA 193 with inner and outer radii as 110$^{\prime\prime}$ and 220$^{\prime\prime}$, respectively. Ancillary response file is generated using {\tt xrtmkarf}. Finally, source spectrum is binned to have at least 20 counts per bin using {\tt grppha}. The Galactic neutral hydrogen column density ($N_{\rm H}=2.73\times10^{21}$ cm$^{-2}$) is taken from \citet[][]{2005AA...440..775K} and we perform the spectral fitting using XSPEC \citep[][]{1996ASPC..101...17A}.

\subsection{Optical-UV Analysis}
Individual \swift~UltraViolet Optical Telescope \citep[UVOT;][]{2005SSRv..120...95R} snapshots are first combined with the task {\tt uvotimsum} and source magnitudes are extracted using {\tt uvotsource}. For the latter, we consider a source region of 5$^{\prime\prime}$ radius and the background is chosen from a nearby region of 30$^{\prime\prime}$ free from source contamination. We de-absorb the magnitudes for Galactic extinction following \citep[][]{2011ApJ...737..103S} and convert them to energy flux units using the zero points provided in \citet[][]{2011AIPC.1358..373B}.

The NOT observations were made using the ALFOSC (Andalucia Faint Object Spectrograph and Camera) instrument at four epochs between MJD 58128$-$58138 (2018 January 10$-$20). Standard IRAF\footnote{http://iraf.noao.edu/} data reduction (de-biasing and flat field correction with twilight flats) is carried out and differential photometry is performed. The brightness is estimated against five SDSS $r'$ stars between 15.3 and 17.4 magnitude. The standard deviation of the zero point is measured as 0.02 mag.

Optical polarimetric observations of DA 193 were taken on MJD 58130$-$58131 (2018 January 12$-$13) from SPOL CCD Imaging/Spectropolarimeter attached to 1.54 m Kupier telescope. The details of the data reduction procedure can be found in \citet[][]{2009arXiv0912.3621S} and here we directly use the publicly available $V$-band polarization data.

\section{Results and Discussion}{\label{sec4}}
\subsection{A New \gm-ray Emitting Blazar}\label{subsec:gm}
\begin{figure*}[t!]
\hbox{
\includegraphics[scale=0.5]{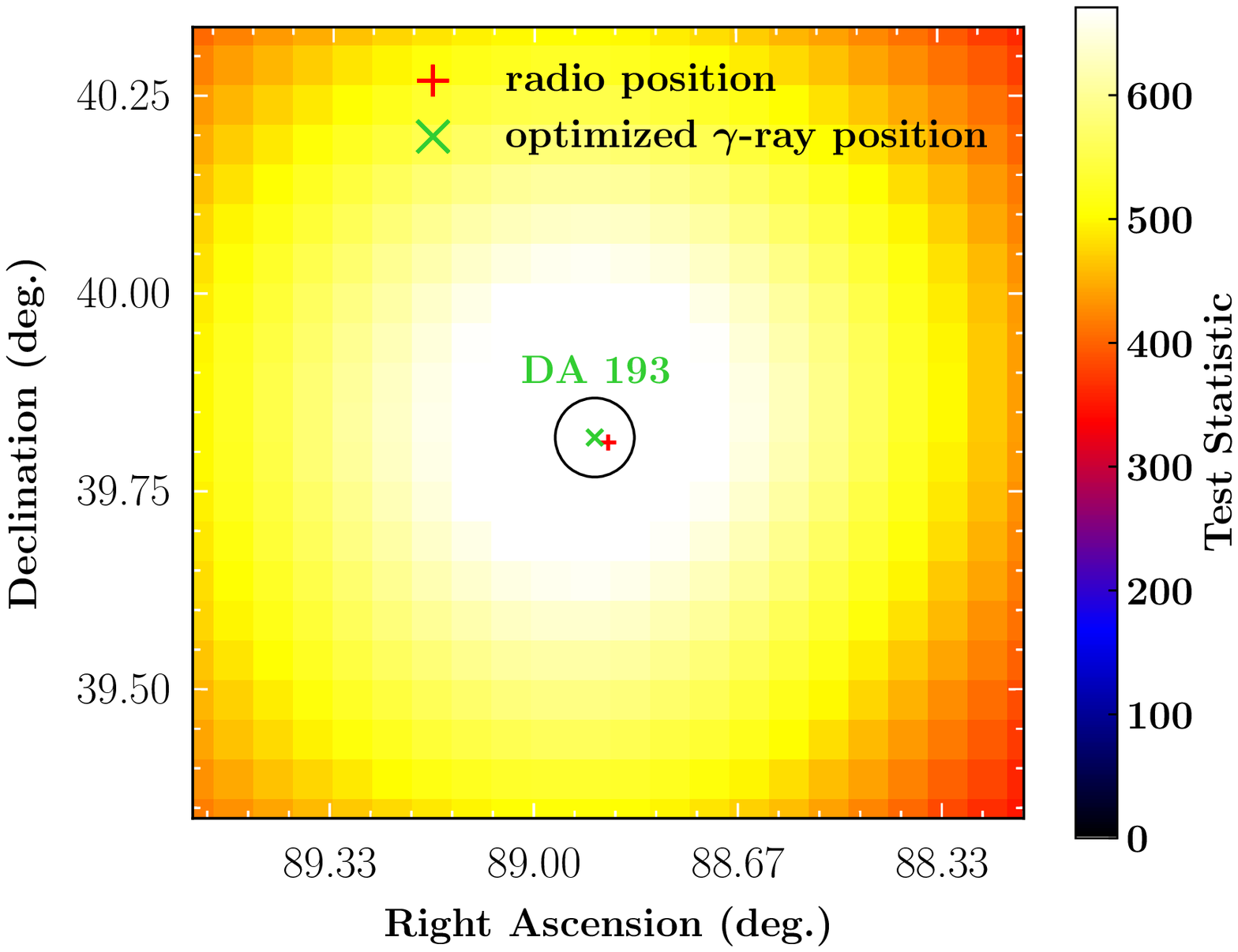}
\hspace{1.0cm}
\includegraphics[scale=0.27]{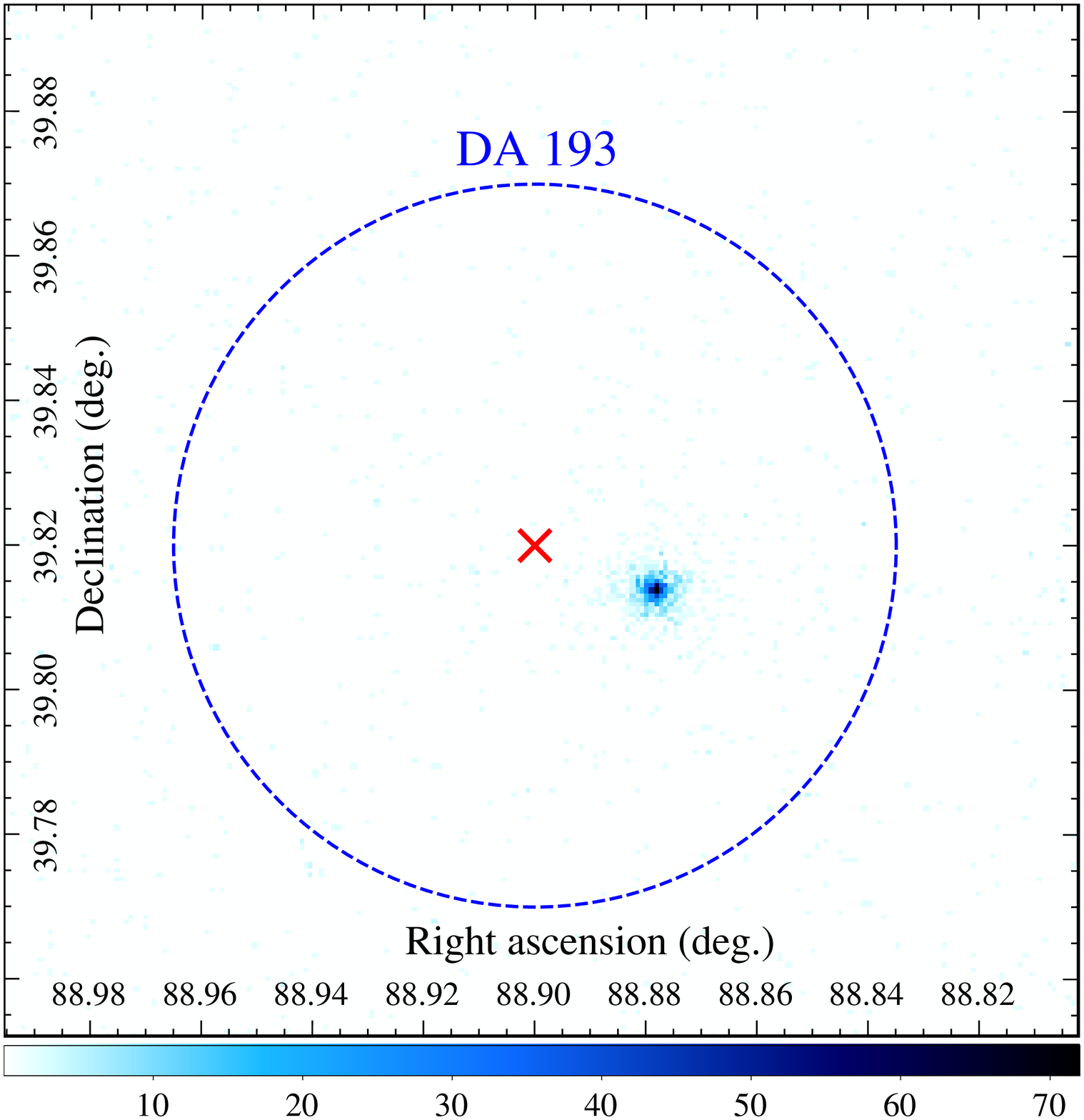}
}
\caption{Left: 0.1$-$300 GeV test statistic map of DA 193, generated for the period MJD 54683$-$58137. Image scale is 0.05 degree per pixel and the black circle denotes the 95\% positional uncertainty derived from the \fermi-LAT data analysis. The radio and optimized \gm-ray positions are also shown, as labelled. Right: This panel shows the X-ray counts map of the DA 193 field generated by combining all \swift-XRT pointings. Blue dashed circle denotes the 95\% error uncertainty in the optimized \gm-ray position which is shown with the red cross. The colorbar has the unit of counts per pixel. The spatial position of the bright X-ray source present in the field is consistent with DA 193.\label{fig_tsmap}} 
\end{figure*}

\begin{figure*}
\hbox{
\includegraphics[scale=0.48]{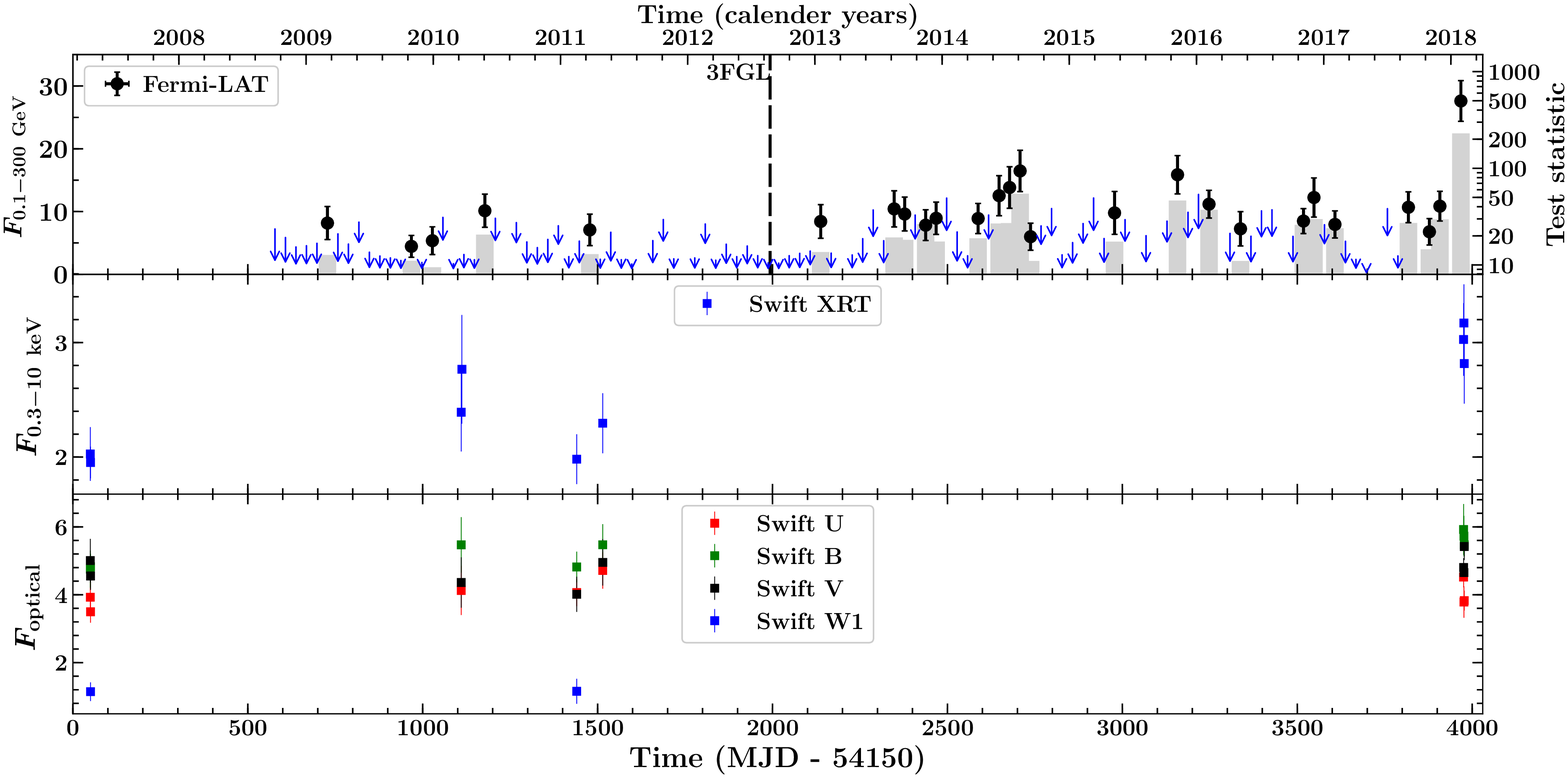}
}
\caption{Top: Monthly binned \gm-ray light curve of DA 193. Black circles represent the source photon flux for time bins with TS$>$9. For TS$<$9,  we derive 2$\sigma$ flux upper limits which are shown with blue downward arrows. Vertical dashed line denotes the time period covered in the 3FGL catalog since the beginning of the \fermi~mission. Light grey bars represent the TS values in bins with TS$>$9. Middle and Bottom: \swift-XRT and UVOT observations of DA 193 with each data point corresponds to one observation id. The rightmost points denote three \swift~pointings taken during our ToO campaign in 2018 January. \fermi-LAT data points are in units of 10$^{-8}$ \phflux, whereas, both XRT and UVOT data have the units of 10$^{-12}$ \ergflux. \label{fig_lc1}} 
\end{figure*}

DA 193 is not present in any of the published \fermi-LAT catalogs. This suggests that either it is a \gm-ray faint blazar or it remained in quiescence for a long time. Contrarily, it is bright in hard X-rays \citep[][]{2018ApJS..235....4O} which indicates it to be a typical high-redshift MeV blazar. 

We search for \gm-ray emission from DA 193 using Pass 8 dataset covering more than 9 years of the \fermi-LAT operation. We start by considering only 3FGL sources within the ROI and generate TS maps to search unmodeled \gm-ray sources that have TS$>$25. A \gm-ray bright source with TS = 671 is identified whose optimized position (corresponding to the maximum of the TS map, R.A.: 05$^h$55$^m$ 36.1$^s$, J2000; Decl.: +39$^{d}$49$^{\prime}$12.0$^{\prime\prime}$, J2000) is found to be spatially consistent with the radio position of DA 193.  The angular separation between the radio and the optimized \gm-ray coordinates is derived as 0$^{\circ}$.02 while the 95\% error radius of the optimized \gm-ray position is 0$^{\circ}$.05. This suggests a positional association of DA 193 with the \gm-ray source (see the left panel of Figure \ref{fig_tsmap}). We also generate an X-ray counts map by combining all of the \swift-XRT pointings of DA 193 and show it in the right panel of Figure \ref{fig_tsmap}. As can be seen, there is only one bright X-ray source present within the 95\% error circle, which is DA 193. This observation provides further evidence of DA 193 being a new blazar in the \gm-ray sky\footnote{A \gm-ray source positionally consistent with DA 193 is present in the recently released FL8Y with name FL8Y J0555.6+3948.}. 

The \gm-ray spectral parameters of DA 193 are as follows: 0.1$-$300 GeV flux = $(5.2 \pm 0.6)\times10^{-8}$ \phflux, photon index = $2.9\pm0.1$. These are similar to those typically observed from luminous high-redshift blazars \citep[][]{2016ApJ...825...74P,2017ApJ...837L...5A}. Its 0.1$-$300 GeV spectrum is very soft and, together with a relatively flat X-ray spectrum as observed from the \swift-BAT, this implies the IC peak to be located at $\sim$MeV energies. Furthermore, we also searched for curvature in the \gm-ray spectrum. This is done by fitting a log-parabola model and comparing the likelihood value ($\mathcal{L_{\rm LP}}$) with that obtained from the power-law fit ($\mathcal{L_{\rm PL}}$). Following \citet[][]{2012ApJS..199...31N}, we compute TS of the curvature: $TS_{\rm curve}=2(\log\mathcal{L_{\rm LP}}-\log\mathcal{L_{\rm PL}})$. The derived $TS_{\rm curve}$ is $<$1 and therefore no significant curvature is found.

Using the optimized \gm-ray coordinates derived above, we also performed the data analysis for the first four years of the \fermi~mission, i.e., time period covered in the 3FGL catalog. This exercise reveals a TS = 21, which is less than the threshold (25) set in the LAT catalogs and thus explains the reason of DA 193 not included in them.

\subsection{\gm-ray Variability}
\begin{figure*}
\hbox{
\includegraphics[scale=0.5]{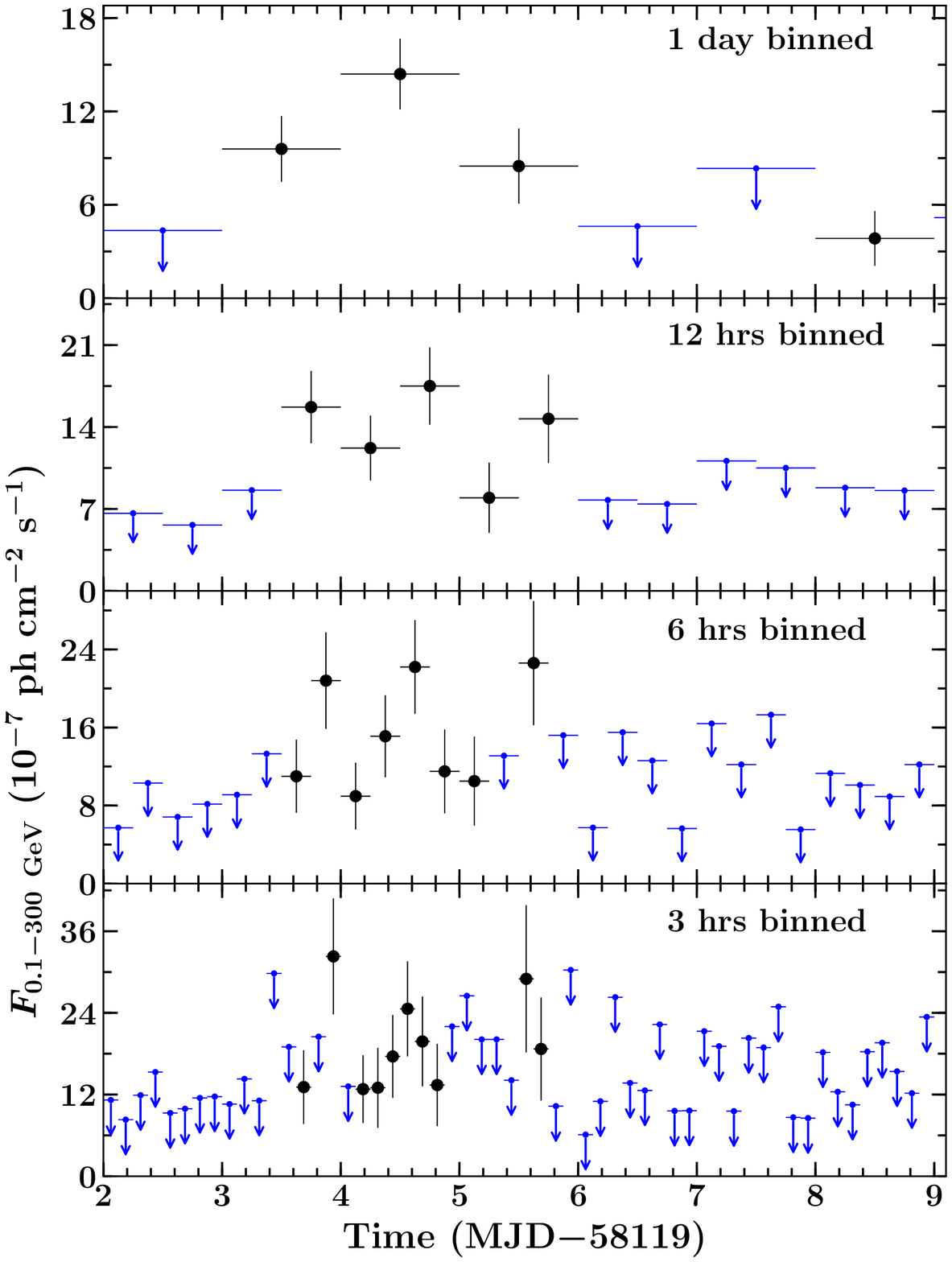}
\includegraphics[scale=0.5]{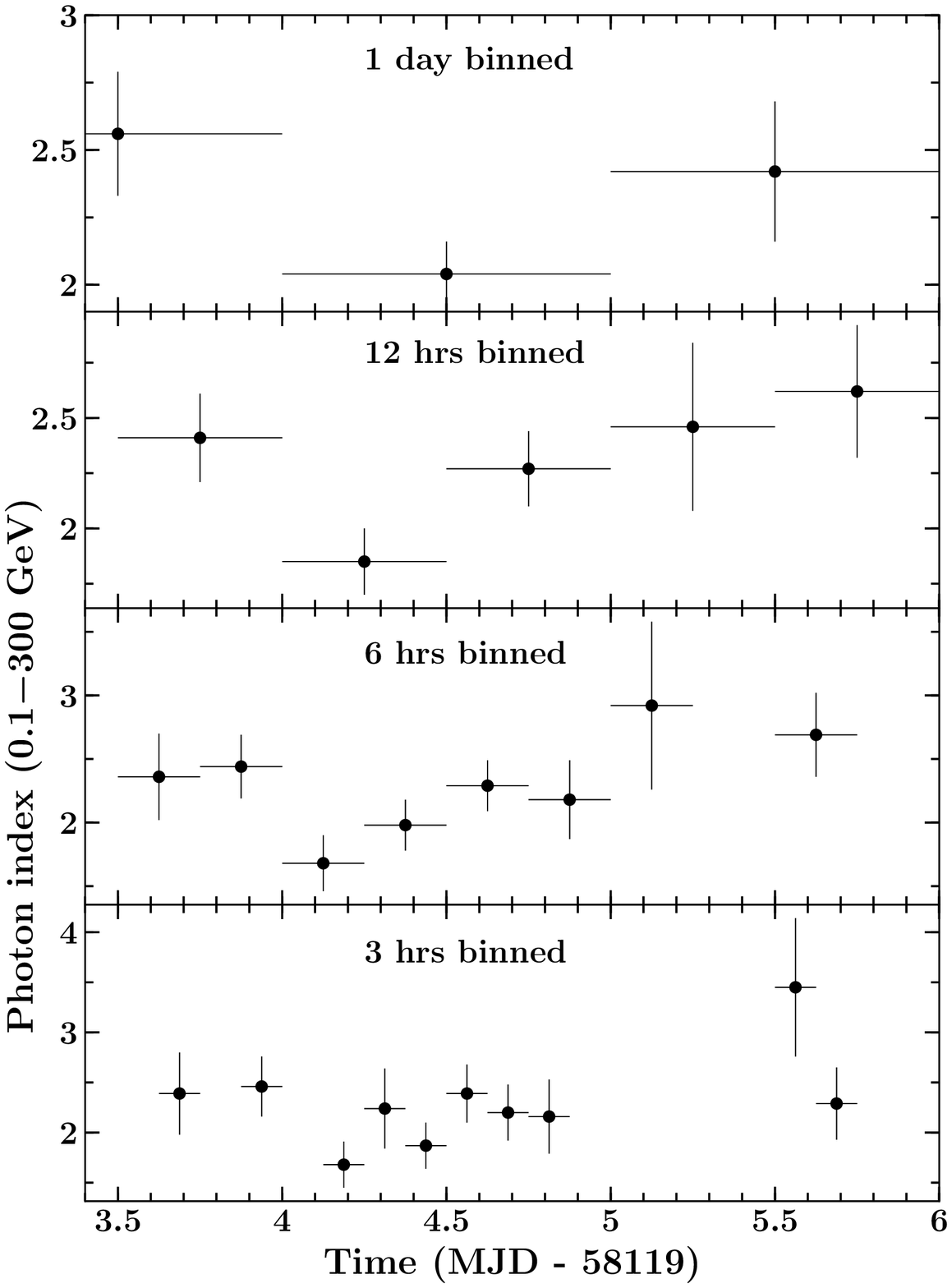}
}
\caption{Left: The \gm-ray light curve of DA 193 focusing on the flaring activity period. Adopted time bins are labelled in different panels. Integrated photon flux upper limits are computed at 2$\sigma$ confidence level. Right: Variation of the \gm-ray photon index during the period of the high activity. MJD 58119 correspond to 2018 January 1.\label{fig_lc2}} 
\end{figure*}

Once we establish the identity of DA 193 as a \gm-ray emitting blazar, we explore its temporal behavior in the \gm-ray band. This is done by generating monthly-binned light curve, which is shown in Figure \ref{fig_lc1}. We find hints of low-level activity during the time period covered in the 3FGL catalog, however, in none of the time bins the source detection is found to be $>$5$\sigma$ significant. There are some moderate level temporal flux variations after that, mainly around MJD 56880 (2014 August). The source entered in its brightest \gm-ray flux state (till now) in the last time bin, i.e. at the beginning of 2018 January.

DA 193 became active in the first week of 2018 and its daily binned \gm-ray flux exceeded 1$\times$10$^{-6}$ \phflux~on MJD 58123 (2018 January 5). Interestingly, the \gm-ray spectrum is found to be extremely hard with photon index $<$2, i.e. a rising spectrum in $\nu F_{\nu}$ vs. $\nu$ SED notation \citep[][]{2018ATel11137....1A}. To study this peculiar event in detail, we focus on the time-period MJD 58122$-$58128, i.e. 2018 January 4$-$10. We generate \gm-ray light curves in various time binning as shown in the left panel of Figure \ref{fig_lc2} and scan them to search for rapid flux variations, however, find none. 

The brightest \gm-ray flux, estimated using the 3-hr binned light curve, is $F_{\gamma}=(3.2\pm0.9)\times 10^{-6}$ \phflux~in the bin centered at MJD 58122.937. The associated photon index is noted as $2.5\pm0.3$ with TS = 49. This implies an isotropic \gm-ray luminosity $L_{\gamma}=(1.3\pm0.4)\times10^{50}$ \lum. Though the observed $L_{\gamma}$ is less than the maximum ever observed from a non-gravitationally-lensed blazar \citep[3C 454.3, $L_{\gamma}\approx2\times10^{50}$ \lum;][]{2011ApJ...733L..26A}, a different scenario emerges when we compare the emitted powers in the jet-frame.  \citet[][]{2014PASJ...66...92L} reported a bulk Lorentz factor $\Gamma_{\rm b}=22$ by modeling the SED of 3C 454.3 during the flaring period in 2010 November \citep[see also,][]{2012ApJ...758...72W}. This, in turn, gives $L_{\rm jet}=L_{\gamma}/2\Gamma_{\rm b}^2\simeq2\times10^{47}$ \lum. For DA 193, on the other hand, the power emitted during the flare is $L_{\rm jet}\simeq5\times10^{47}$ \lum, assuming $\Gamma_{\rm b}=11$ (Section \ref{subsec_sed}), which is about two-times larger than that observed from 3C 454.3. \citet[][]{2014MNRAS.444.3040O} studied the \gm-ray outburst of another high-redshift blazar TXS 0536+145 ($z=2.69$) and reported $L_{\gamma}=6.6\times10^{49}$ \lum~and $\Gamma_{\rm b}=30$. In addition to that, $L_{\gamma}=1.6\times10^{50}$ \lum~and $\Gamma_{\rm b}=19$ was reported for the 2011 December \gm-ray flare of the bright high-redshift blazar S5 0836+71 \citep[$z=2.17$,][]{2015ApJ...804...74P}. In the jet-frame, this gives $L_{\rm jet}\simeq4\times10^{46}$ \lum~and $2\times10^{47}$ \lum, respectively for TXS 0536+145 and S5 0836+71, which are smaller than that found for DA 193. Therefore, the \gm-ray flaring activity of DA 193 was intrinsically more powerful than the brightest \gm-ray flare from 3C 454.3 or probably other high-redshift blazars studied so far.

The right panel of Figure \ref{fig_lc2} represents the spectral behavior of the source during the flaring episode. As can be seen, the \gm-ray spectrum significantly ($>$5$\sigma$) hardens during the flare. In particular, a photon index of $1.7\pm0.2$ was noted on MJD 58123.187, with $F_{\gamma}=(1.3\pm0.5)\times10^{-6}$ \phflux~and TS = 67. Such a hard \gm-ray spectrum is typically observed from high-synchrotron peaked BL Lac objects \citep[e.g.,][]{2015ApJ...810...14A} and, to our knowledge, has never been observed from a $z>2$ blazar.

The detection of the GeV flare from a high-redshift blazar is a rare phenomenon due to these objects being faint in the \gm-ray band. What makes this peculiar flaring activity even rarer is the observation of an extremely hard \gm-ray spectrum, as described above. This is due to two reasons. First is the $K$-correction: for increasing redshifts \citep[][]{2002astro.ph.10394H}, which makes the IC peak to be located closer to the hard X-ray band and hence brighter with respect to \fermi-LAT energy range. The shift of the SED peaks to lower frequencies with the increase in the bolometric luminosity, is another effect which makes high-redshift blazar fainter with steeper spectrum in \gm-rays.

With the motivation to perform a comparative study of different activity states of DA 193, we define two time-periods. MJD 54683$-$58118 (2008 August 8 to 2017 December 31) is adopted as a low activity period and MJD 58122$-$58128 (2018 January 4$-$10) is assumed to cover the flaring episode. We note that the adopted low activity duration may not represent the true quiescence of the blazar since a few sparse moderate level flux enhancements are observed (Figure \ref{fig_lc1}). However, a true quiescence may not serve the purpose as we need to generate a meaningful \gm-ray SED. Therefore, we can call MJD 54683$-$58118 as a low activity period with respect to the GeV flare observed in 2018 January. The spectral parameters derived from the \fermi-LAT data analysis for these two periods are presented in Table \ref{tab:sed_par}.

\subsection{X-ray Analysis}
\begin{figure}
\hbox{
\includegraphics[width=\columnwidth]{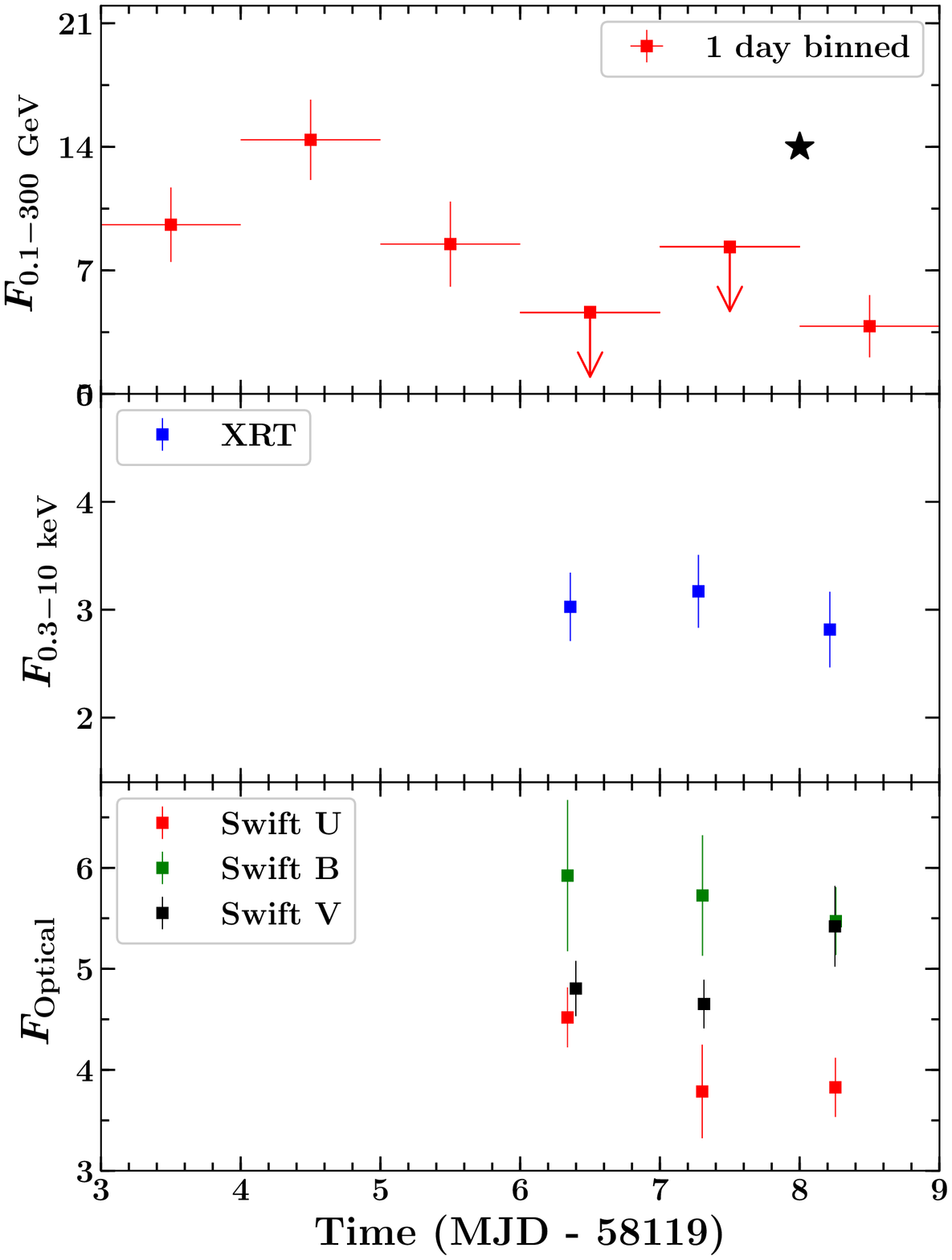}
}
\caption{Multi-wavelength light curve showing the \fermi-LAT and \swift~observations of DA 193 during GeV flare. \fermi-LAT data points are in units of 10$^{-7}$ \phflux, whereas, \swift-XRT and UVOT fluxes are in units of 10$^{-12}$ erg cm$^{-2}$ s$^{-1}$. Black star in the top panel represent the time of \nustar~ToO observation.\label{fig_lc3}} 
\end{figure}

In the middle panel of Figure \ref{fig_lc3}, we show 0.3$-$10 keV count rate measured from the three \swift-XRT observations taken during the GeV flare. Comparing the X-ray light curve with the \gm-ray one (top panel), we find that we were able to follow the source during the decaying phase of the flare. Moreover, no significant flux variability is observed and therefore, we combine all of the three \swift-XRT observations to improve the photon statistics and generate a 0.3$-$10 keV spectrum representing the flaring period. Following a similar approach, we add all of the previous \swift-XRT measurements to generate an X-ray SED corresponding to the low activity state (see the middle panel in Figure~\ref{fig_lc1}). We fit an absorbed power-law model to derive the spectral parameters for both X-ray SEDs and provide them in Table~\ref{tab:sed_par}.

A simple power-law model explains the \nustar~spectrum reasonably well and the associated spectral parameters are shown in Table \ref{tab:sed_par}.

\begin{figure}[t!]
\hbox{\hspace{-0.2cm}
\includegraphics[width=\columnwidth]{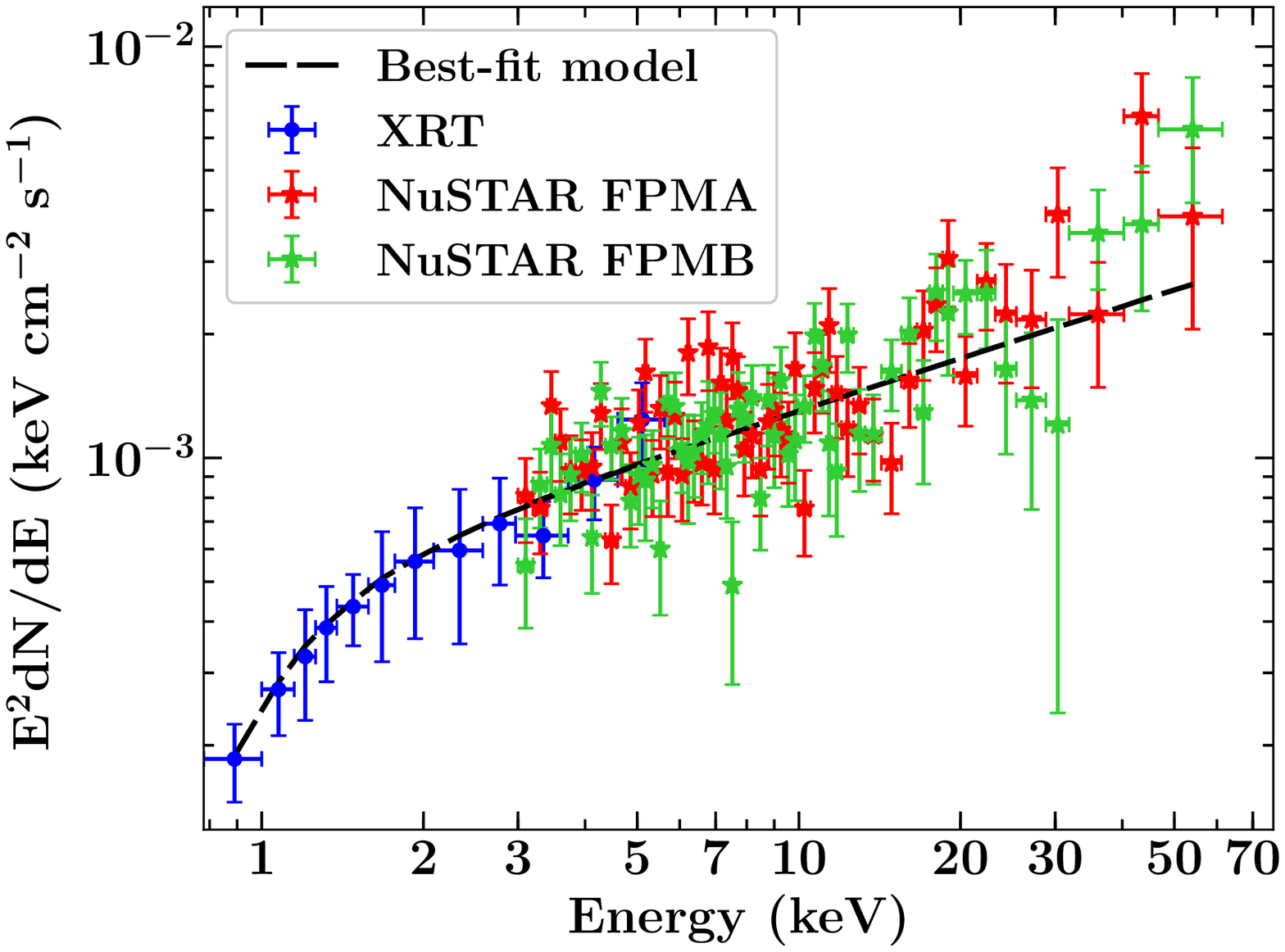}
}
\caption{This plot represent the joint \swift-XRT and \nustar~spectrum of DA 193 taken during the GeV flare. The best-fit absorbed power-law model is shown with the dashed line. The fitting is performed in 0.3$-$79 keV and the joint spectrum is rebinned to have at least 5$\sigma$ significance in each bin. The source is well detected by \nustar~up to 70 keV.\label{fig_nustar}} 
\end{figure}

We also perform a joint \nustar~and \swift-XRT fitting using the data taken during the flaring period and adopt an absorbed power-law model with neutral hydrogen column density fixed to the Galactic value. An inter-calibration constant is considered while performing the fit, to allow for the cross-calibration uncertainties between \nustar~(FPMA and FPMB) and \swift. This constant is fixed to unity for FPMA and left free to vary for FPMB and \swift-XRT spectra. It is found to be 0.98$\pm$0.04 and 0.91$\pm$0.10 for FPMB and \swift-XRT, respectively. The spectral parameters are provided in Table \ref{tab:sed_par} and the spectra, along with the best-fitted model, are shown in Figure \ref{fig_nustar}.  Since high-redshift blazars typically exhibit a significant soft X-ray spectral break \citep[e.g.,][]{2001MNRAS.323..373F,2004MNRAS.350L..67W}, we also test an absorbed broken power-law model ($\chi^2$/dof = 138/118), however, do not find a significant improvement with respect to an absorbed power-law model (f-test probability = 0.01).

\subsection{Optical-UV Observations}
\begin{figure}[t!]
\hbox{
\includegraphics[width=\columnwidth]{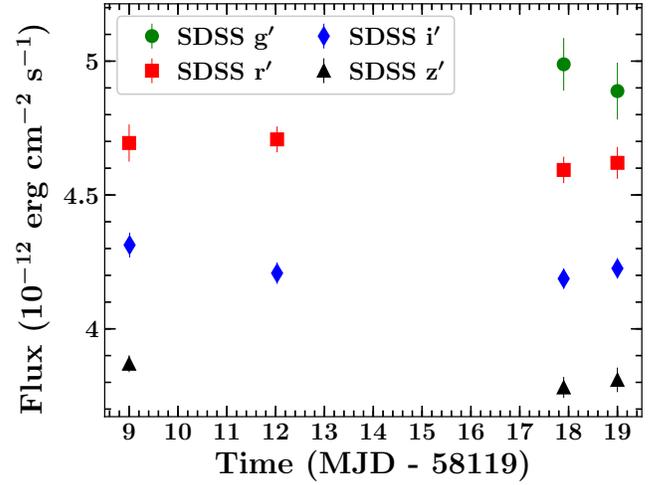}
}
\caption{Multi-band optical light curve of DA 193 taken from Nordic Optical Telescope during the decaying phase of the GeV flare in 2018 January. The filters used are SDSS $g'$, $r'$, $i'$, and $z'$. No significant variability is detected.\label{fig_not}} 
\end{figure}

In the bottom panel of Figure \ref{fig_lc3}, we show the temporal variation of the optical flux in the $V$, $B$, $U$, and $W1$ bands as observed from the \swift-UVOT. The source, on the other hand, is not detected in $M2$ and $W2$ filters. As evident from this plot, DA 193 did not show any significant flux variability during the \gm-ray flaring period. Similar results are derived from optical photometric observations taken from NOT (see Figure \ref{fig_not}). The lack of the optical flux variability can be explained on the basis of two facts: either the source returned to faint state and/or the optical-UV emission is dominated by thermal radiation from the accretion disk which is not expected to vary on short timescales. In fact, our SED modeling results confirms that the IR-UV spectrum of DA 193 is indeed dominated by the accretion disk emission (see below). Furthermore, on the nights of MJD 58130 and 58131 (2018 January 12$-$13), Steward observatory detected $V$-band polarization of 2.33$\pm$0.25\% and 1.66$\pm$0.19\%, respectively. The associated optical polarization angles are 142.7$\pm$3.1 degree and 165.2$\pm$3.3 degree, respectively.  Such a low degree of the optical polarization further provides supportive evidences for the accretion disk origin of the optical radiation. There were a few multi-wavelength campaigns where a significant optical-\gm-ray correlation and a systematic rotation of the optical polarization angle were noted during the \gm-ray flaring episodes \citep[e.g.,][]{2008Natur.452..966M,2010Natur.463..919A,2013ApJ...768...40L}. Unlike them, we do not have a dense sampling of the optical flux and polarization measurement and hence a definitive conclusion about the origin of the optical emission and also a connection between the optical/\gm-ray activity and the jet and accretion disk cannot be made. As can be seen in the bottom right panel of Figure \ref{fig_lc1}, previous \swift-UVOT observations do not show any significant variability and therefore we combine them to generate low activity state optical SED.

\subsection{Black Hole Mass and Accretion Disk Luminosity}\label{subsec:bh}

\begin{figure}
\hbox{
\includegraphics[width=\columnwidth]{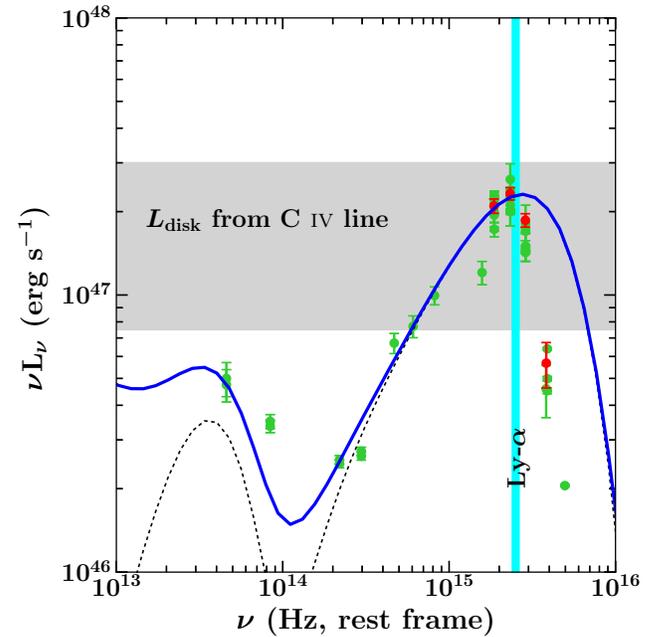}
}
\caption{Rest-frame IR-UV SED of DA 193 along with the standard accretion disk model. Only data points redward to the Lyman-$\alpha$ frequency (vertical cyan line) are considered for the modeling. Black dashed line shows the thermal radiation from the torus and the accretion disk. Blue solid line is the total emission considering also the non-thermal jetted radiation. The shaded region corresponds to the peak disk luminosity derived from the C~{\sc iv} emission line parameters and we assume an uncertainty of 0.3 dex. See text for details.\label{fig_disk}} 
\end{figure}

The central black hole mass can be calculated from the single epoch optical spectroscopic emission line information, assuming that the  BLR is virialized \citep[e.g.,][]{2011ApJS..194...45S}. We use the C~{\sc iv} emission line parameters provided by \citet[][]{2012RMxAA..48....9T} and adopt empirical relations of \citet[][]{2011ApJS..194...45S}. The black hole mass is derived as $(5.5\pm0.9)\times10^9$ \msun. Moreover, we can derive the BLR luminosity from the emission line luminosity by following the scaling relations of \citet[][]{1991ApJ...373..465F} and \citet[][]{1997MNRAS.286..415C}. The accretion disk luminosity can then be inferred assuming that the BLR reprocesses 10\% of the accretion disk emission. With this approach, we determine the accretion disk luminosity as $(1.3\pm0.1)\times10^{47}$ \lum.

\begin{figure*}
\hbox{
\includegraphics[scale=0.75]{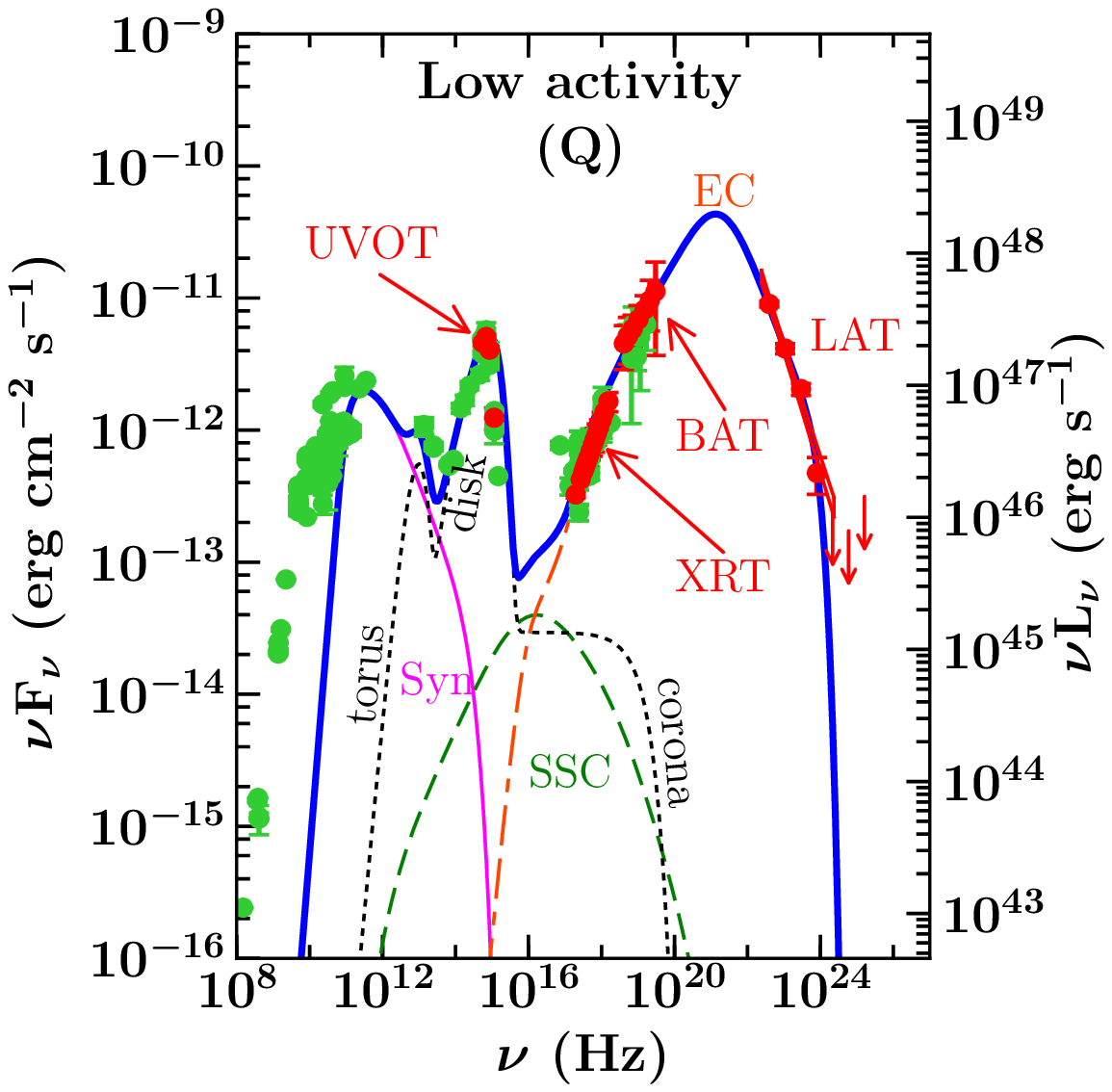}
\includegraphics[scale=0.75]{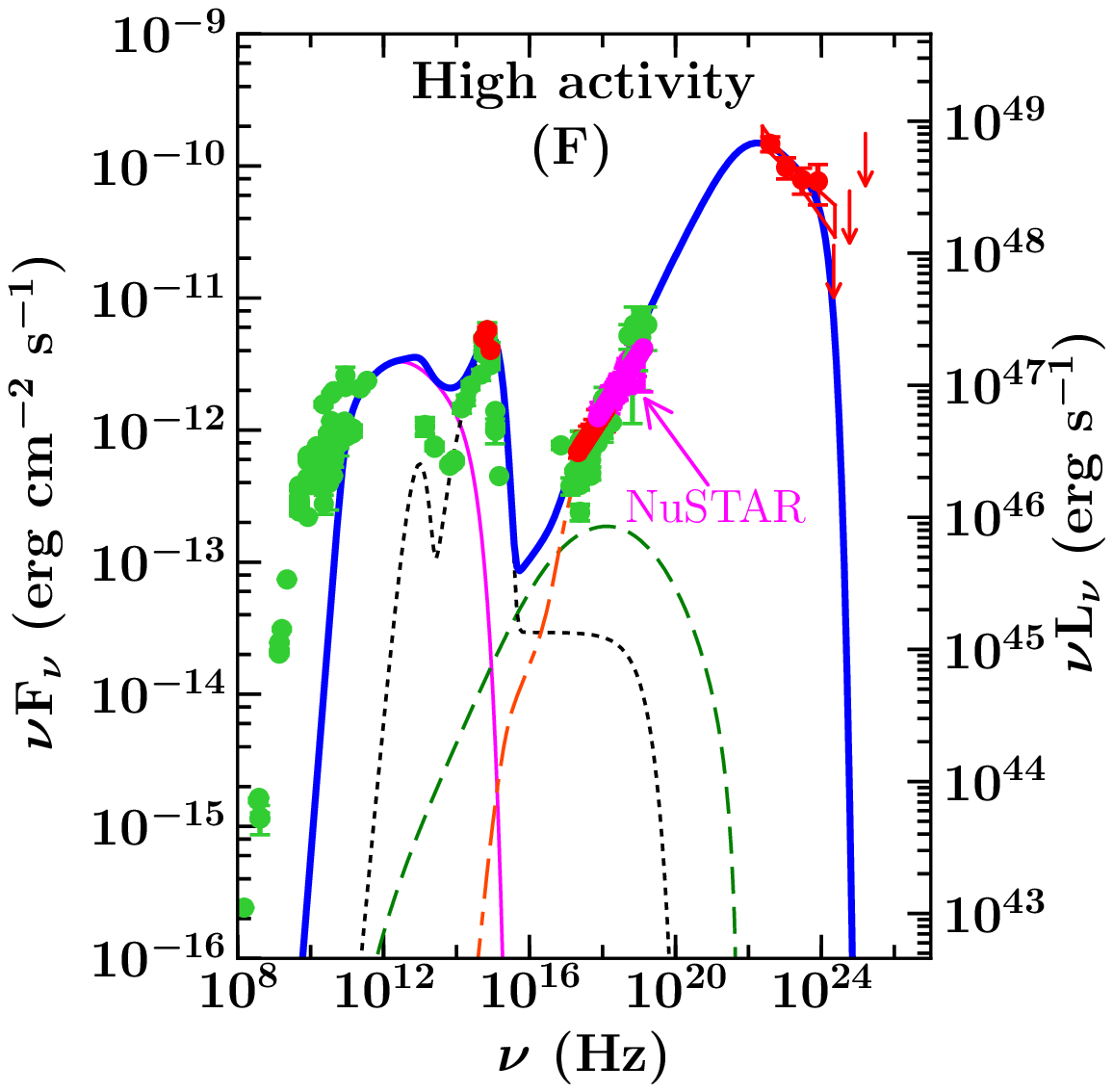}
}
\caption{Left: Broadband SED of DA 193 during the low activity state. The data analyzed by us are shown with red circles, whereas, archival observations are represented by green circles. In the LAT energy range, 2$\sigma$ flux upper limits are denoted by downward arrows. Black dashed line corresponds to thermal emission from the torus, accretion disk, and X-ray corona, as labeled. Pink solid, green dashed, and orange dash-dash-dot lines represent synchrotron, synchrotron self Compton, and external Compton processes, respectively. Blue solid line is the sum of all the radiative components. Right: The elevated state SED of DA 193 generated by averaging the data for the whole flaring period (see, Figure \ref{fig_lc3}). The adopted time-periods to generate both low- and high-activity SEDs are defined in Table \ref{tab:sed_par}. See the text for details.\label{fig_sed}} 
\end{figure*}

Another method to derive the black hole mass and the accretion disk luminosity is by modeling the IR-UV spectrum by a standard optically thick, geometrically thin disk \citep[][]{1973A&A....24..337S}. However, the primary constraint with this technique is that the accretion disk emission (big blue bump) should be visible. When this condition is met, the parameters derived from the disk modeling reasonably agree with that computed from the conventional optical spectroscopic method \citep[e.g.,][]{2017ApJ...851...33P}. The flux variability could affect the parameter estimation, however, it is not expected to be a major factor since the disk emission usually does not vary over short timescales. The IR-UV emission from DA 193 shows a prominent bump which we interpret due to accretion disk and show it in Figure \ref{fig_disk}. To model the observed bump, we do not consider data points blueward of the hydrogen Lyman-$\alpha$ frequency (vertical cyan line) due to possible absorption by intervening Lyman-$\alpha$ clouds whose nature is uncertain.  We find a black hole mass of $3\times10^9$ \msun~and a disk luminosity of $3\times10^{47}$ \lum, which matches within a factor of two and three, respectively, with that determined from the optical spectroscopic method.

\subsection{Spectral Energy Distribution}\label{subsec_sed}

\subsubsection{Modeling}
We generate the broadband SED of DA 193 covering the low and flaring activity states following the procedure outlined in Section \ref{sec3}. We also consider the archival measurements from ASDC SED builder\footnote{https://tools.asdc.asi.it/}. We note that the low activity state SED, which is generated by averaging years of the multi-wavelength data, provides us information about an overall average physical behavior of the blazar and may not represent any specific activity state. For the flaring state, on the other hand, we have used available near-simultaneous observations. DA 193 is not detected above 10 GeV and hence EBL absorption is negligible.

Broadband SEDs are then modeled using a one-zone, leptonic radiation model \citep[see, e.g.,][]{2008ApJ...686..181F,2009ApJ...692...32D,2009MNRAS.397..985G}. We assume a spherical emission region located at a distance $R_{\rm dist}$ from the black hole and move along the jet-axis with a bulk Lorentz factor $\Gamma_{\rm b}$. Considering a uniform but randomly oriented magnetic field, relativistic electrons, following a broken power-law energy distribution, radiate via synchrotron and IC processes. For the latter, we consider low-energy seed photons originated both inside the jet (synchrotron photons) and outside of it, i.e. from the accretion disk, BLR, and torus. The radiation profile of the accretion disk is assumed to follow a multi-color blackbody pattern \citep[][]{2002apa..book.....F}. Both the BLR and the torus are considered as spherical blackbodies peaking at Lyman-$\alpha$ frequencies and a characteristic torus temperature $T_{\rm IR}$, respectively. The jet is assumed to have a conical shape with semi-opening angle of 0.1 rad and the emission region covers the entire cross-section of the jet. We calculate various jet powers following \citet[][]{2008MNRAS.385..283C} and in particular, kinetic jet power is derived by assuming equal number density of electrons and cold protons. The modeled SEDs are shown in Figure \ref{fig_sed} and the parameters are reported in Table \ref{tab:sed}.

\subsubsection{Interpretation}
\begin{figure*}[t!]
\hbox{
\includegraphics[width=\columnwidth]{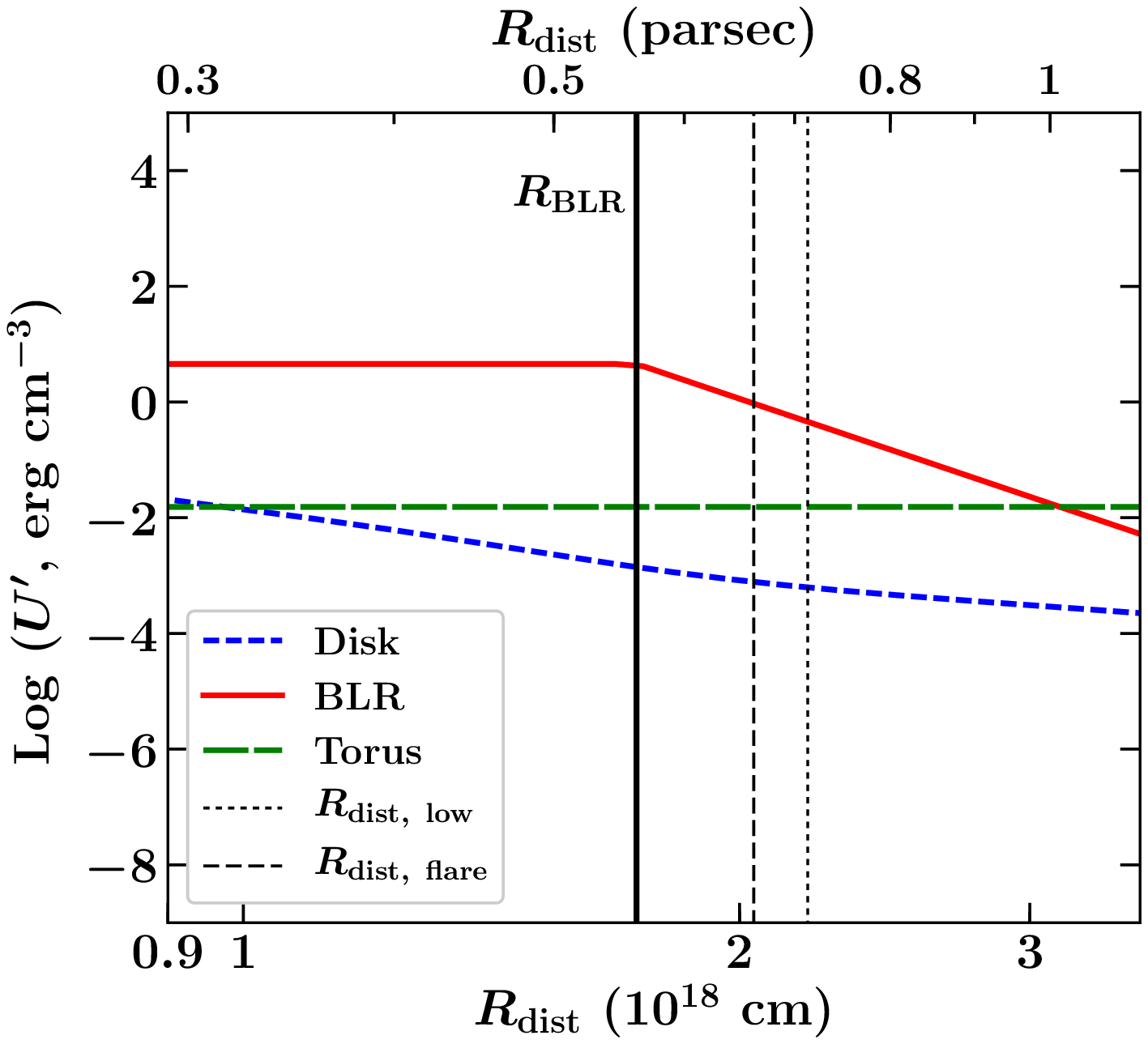}
\includegraphics[width=\columnwidth]{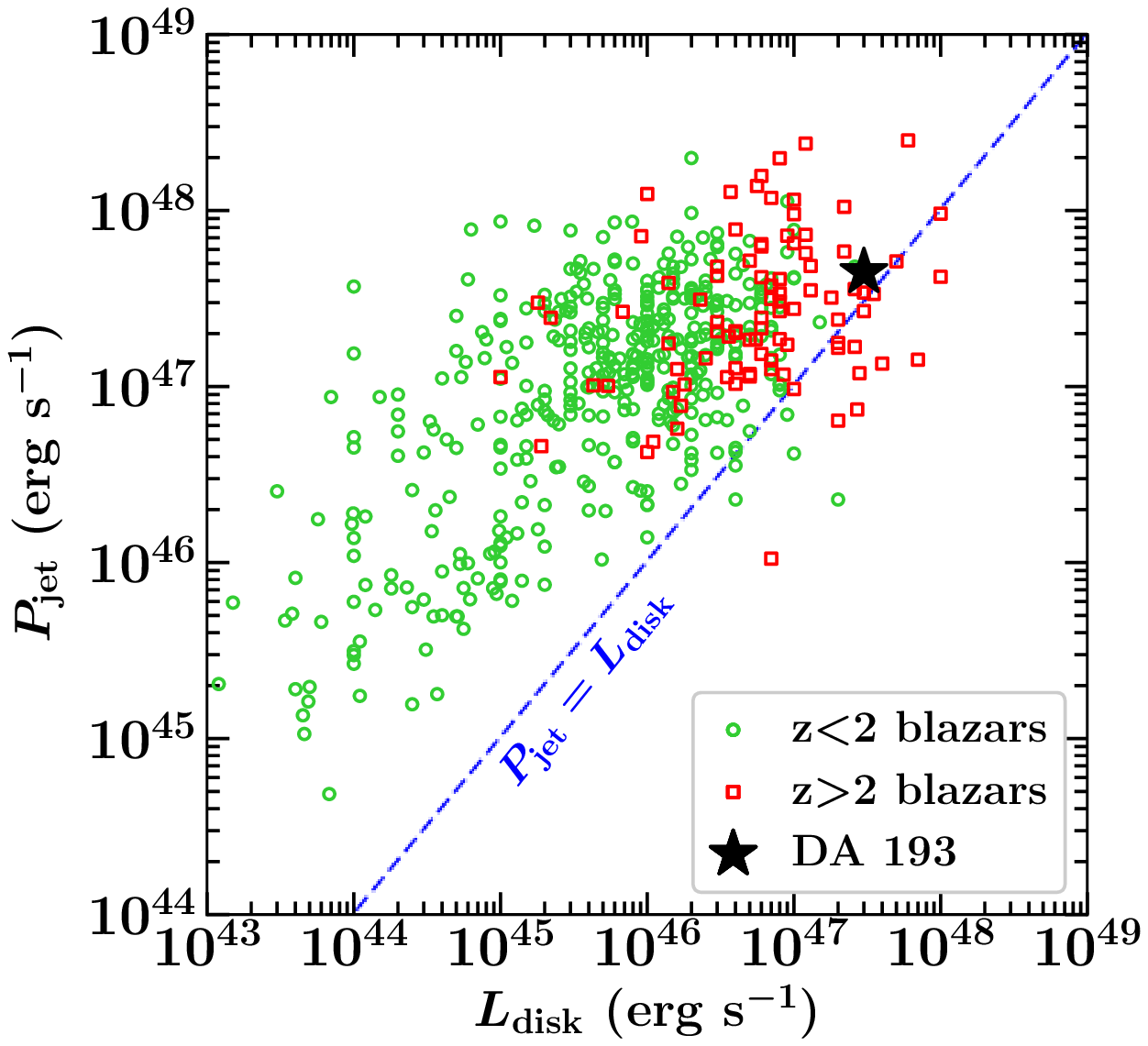}
}
\caption{Left: Variation of the comoving-frame radiative energy densities of various AGN components as a function of the distance from the central black hole, as labelled. Vertical black solid line represents the inner boundary of the BLR, whereas, dotted and dashed lines denote the locations of the emission region during the low and flaring activity states. Right: The jet power as a function of the accretion disk luminosity for a large sample of blazars studied in \citet[][]{2017ApJ...851...33P}. Blazars beyond $z=2$ are shown with red squares while $z<2$ sources are denoted by limegreen circles. The black star reporesents DA 193 with the jet power computed during its low activity state. Blue dashed line corresponds to the equality of the jet power and the accretion luminosity. Note that, the jet power of DA 193 is computed assuming a two-sided jet, for equal comparison. \label{fig_ene_den}} 
\end{figure*}

The low activity SED of DA 193 (Figure \ref{fig_sed}, left panel) resembles well with that typically observed for other high-redshift blazars \citep[e.g.,][]{2015ApJ...807..167T}. The synchrotron spectrum peaks at $\sim$sub-mm wavelengths and the optical-UV emission is dominated by the accretion disk (see also, Figure \ref{fig_disk}). We reproduce entire X-ray to \gm-ray SED with external Compton (EC) mechanism with seed photons primarily originating from the BLR with some contribution from the torus (see also, Figure \ref{fig_ene_den}). A flat X-ray spectrum also hints for the EC origin of the radiation, rather than due to synchrotron self Compton (SSC) process which predicts a relatively softer X-ray spectrum. The requirement to keep SSC below the observed X-ray spectrum, in turn, constrains the magnetic field strength and the size of the emission region. Furthermore, in our SED model, radiative energy densities are a function of $R_{\rm dist}$ \citep[following prescriptions of][]{2009MNRAS.397..985G}, and hence the level of the EC spectrum provides further clues about the location of the \gm-ray emitting region (Figure \ref{fig_ene_den}). The shapes of the X-ray and \gm-ray spectra enable us to reliably determine the slopes of the underlying electron energy distribution before and after the break energy ($\gamma_{\rm bk}$).

Recently, \citet[][]{2018MNRAS.477.4749C} have reported that the \gm-ray spectra of the bright \fermi~blazars do not exhibit a break up to tens of GeV as expected if the seed photons for EC process are mainly provided by the BLR. This finding led them to conclude that the primary mechanism for the \gm-ray production should be EC-torus instead of EC-BLR. DA 193, on the other hand, is not even detected above 5 GeV, thus indicating a possible BLR and/or EBL absorption of higher energy \gm-ray photons. Using the EBL model of \citet[][]{2010ApJ...712..238F}, we find the EBL optical depth to be less than unity even for a 40 GeV photon at $z=2.36$, thus suggesting a negligible EBL absorption. Therefore, non-detection of DA 193 above a few GeV suggests a possible attenuation of higher energy \gm-ray photons by the BLR photon field. This sets the location of the emission region close to the BLR which is also supported by our SED modeling results. Another important consequence of considering the EC-torus as the primary mechanism for the \gm-ray production is the location of the high-energy peak. This is because, in Thomson regime, $\nu_{\rm EC}\propto \gamma^2_{\rm bk}\nu_{\rm seed}$, where $\nu_{\rm EC}$ is the EC peak frequency and $\nu_{\rm seed}$ is the characteristic frequency of the seed photon field. Since $\gamma_{\rm bk}$ has a rather low value, constrained from the synchrotron peak location, and the fact that the characteristic energy of BLR photons ($\sim$10 eV) is larger compared to torus ones ($\sim$0.1 eV), the \swift-BAT and \fermi-LAT data enable us to accurately constrain the inverse Compton peak which supports the EC-BLR as the primary mechanism (see also, Figure \ref{fig_ene_den}). Finally, the results derived by modeling a sample of high-redshift blazars \citep[][]{2010MNRAS.405..387G} also indicates the BLR as a primary reservoir for the EC process and our findings are consistent with them.

We reproduce the elevated activity state SED (top right panel in Figure \ref{fig_sed}) by trying to change the minimum number of parameters with respect to that derived for the low state. We notice a major change in the behavior of the electron population. During the GeV flare, not only the \gm-ray flux level increases but also the spectrum significantly hardens. This indicates the hardening of the electron energy distribution after $\gamma_{\rm bk}$ which itself shifts to higher energies (Table \ref{tab:sed}), possibly due to the injection of fresh, highly energetic electrons in the emission region. Interestingly, compared to the low state SED, there are negligible variations at the optical and X-ray frequencies. The optical spectrum shows a bump indicating the dominance of the accretion disk emission. The lack of substantial temporal variations observed from UVOT and NOT (Figure \ref{fig_lc3} and \ref{fig_not}) supports this finding. Considering the X-ray spectrum taken by \swift-XRT alone, we find a hint of softening during the flare (Table \ref{tab:sed_par}), however, due to small exposure and hence low photon statistics, a strong claim cannot be made. On the other hand, the photon index derived from the joint \nustar~and \swift-XRT spectral fitting is compatible to that noted by \citet[][]{2018ApJS..235....4O} from the \swift-BAT observations. Therefore, it can be concluded that there is no significant temporal or spectral evolution in the X-ray band. 

A decrease in the dissipation distance, hence an enhanced BLR radiation field in the comoving frame (Figure \ref{fig_ene_den}, left panel), explains the detection of a Compton dominated SED during the GeV flare. Though the optical emission originates primarily from the accretion disk, our modeling suggests a substantial synchrotron flux enhancement in the IR band. This is due to a shift in the SED peaks to account for the \gm-ray hardening and elevated synchrotron radiation. Finally, as discussed above, a large separation between the synchrotron and EC peaks also indicates the BLR as a primary reservoir of the seed photons for IC scattering. 

In the right panel of Figure \ref{fig_ene_den}, we show the variation of the total jet power ($P_{\rm jet} = P_{\rm e}+P_{\rm B}+P_{\rm p}$) as a function of the accretion luminosity for CGRaBS blazars \citep[][]{2008ApJS..175...97H} whose broadband properties are studied in \citet[][]{2017ApJ...851...33P}. One can immediately notice that high-redshift blazars lie at the highest end of the jet-disk luminosity correlation \citep[see also,][for earlier results]{2010MNRAS.405..387G}. This diagram also suggests that DA 193 is among the systems hosting some of the most powerful relativistic jets and accretion disks.

Comparing various jet powers derived from the low activity state modeling, we find that a major fraction of the total jet power is carried by cold protons in the form of the kinetic power (Table \ref{tab:sed}). Both magnetic and radiative jet powers are about an order of magnitude lower which indicates a low magnetization of the emission region \citep[e.g.,][]{2015MNRAS.451..927Z} and that only a fraction of the kinetic power gets converted to radiation. During the elevated state, on the other hand, we find the radiative power to be almost equal to the power in protons, which hints a high radiative efficiency of the jet during the flaring episode. Blazars, however, are known to spend only a tiny fraction of time in the high activity state \citep[$\sim$1\%, e.g.,][]{2010MNRAS.405L..94T}, therefore, such flaring events (with $P_{\rm r}\sim P_{\rm p}$) are short-lived.

\section{Summary}{\label{sec5}}
We have performed a detailed multi-frequency study of the high-redshift blazar DA 193. Our findings are summarized below.  
\begin{enumerate}
\item We report the first time detection of a significant \gm-ray emission from DA 193.
\item Its long time-averaged \gm-ray spectrum is soft ($\Gamma_{\rm 0.1-300~GeV}$ = $2.9\pm0.1$), whereas, it exhibits a relatively flat hard X-ray spectrum ($\Gamma_{\rm 14-195~keV}$ = $1.5\pm0.4$). These observations are aligned with that typically observed from high-redshift blazars.
\item In the first week of 2018, DA 193 underwent an extremely luminous GeV flare ($L_{\gamma}=(1.3\pm0.4)\times10^{50}$ \lum). The observation of a hard \gm-ray spectrum ($\Gamma_{\rm 0.1-300~GeV}$ = $1.7\pm0.2$) makes this \gm-ray outburst a rare phenomenon. Moreover, the multi-wavelength light curve covering the flaring period suggests that we were able to follow the source during the decaying phase of the flare.
\item The broadband SED of DA 193, during low activity state, is found to be similar to that observed from high-redshift blazars. The optical emission remains accretion disk dominated even during the GeV flare which is further supported from the observation of the low-degree of the optical polarization noted from the Steward observatory.
\item A comparison of the near-simultaneous observations both from space and ground telescopes, i.e. \swift, \nustar~and NOT, covering the flaring period, with previous archival measurements, reveals an insignificant flux change in the optical-X-ray energy range. This probably indicates that the source returned back to quiescence even before the beginning of the multi-wavelength campaign.
\item A simple one-zone radiation model successfully reproduces the multi-frequency observations. According to our analysis, a change in the behavior of the underlying electrons population could be responsible for the observed \gm-ray flare.
\end{enumerate}

We conclude that a continuous monitoring of the \gm-ray sky by the \fermi-LAT is crucial to hunt down the elusive population of the most powerful blazars, similar to DA 193. The \fermi-LAT also enables a unique opportunity to study different flavors of the \gm-ray flares which will improve our understanding of the radiative processes powering the relativistic jets of blazars.

\acknowledgments

Thanks are due to anonymous referees for useful suggestions on the manuscript. We are grateful to \nustar~and \swift~PIs for approving the ToO observations and to the mission operation teams for quickly executing them. Thanks are also due to P. Smith (Steward observatory) for considering our request of observations. VSP acknowledges funding under NASA contracts 80NSSC17K0310 and 80NSSC18K0580. RA is a member of the International Max Planck Research School (IMPRS) for Astronomy and Astrophysics at the Universities of Bonn and Cologne.

The \textit{Fermi} LAT Collaboration acknowledges generous ongoing support from a number of agencies and institutes that have supported both the development and the operation of the LAT as well as scientific data analysis. These include the National Aeronautics and Space Administration and the Department of Energy in the United States, the Commissariat \`a l'Energie Atomique and the Centre National de la Recherche Scientifique / Institut National de Physique
Nucl\'eaire et de Physique des Particules in France, the Agenzia Spaziale Italiana and the Istituto Nazionale di Fisica Nucleare in Italy, the Ministry of Education, Culture, Sports, Science and Technology (MEXT), High Energy Accelerator Research Organization (KEK) and Japan Aerospace Exploration Agency (JAXA) in Japan, and the K.~A.~Wallenberg Foundation, the Swedish Research Council and the Swedish National Space Board in Sweden. Additional support for science analysis during the operations phase is gratefully acknowledged from the Istituto Nazionale di Astrofisica in Italy and the Centre National d'\'Etudes Spatiales in France. This work performed in part under DOE Contract DE- AC02-76SF00515.

This work made use of data from the \nustar~mission, a project led by the California Institute of Technology, managed by the Jet Propulsion Laboratory, and funded by the National Aeronautics and Space Administration. We thank the \nustar~Operations, Software, and Calibration teams for support with the execution and analysis of these observations. This research has made use of the \nustar~Data Analysis Software (NuSTARDAS) jointly developed by the ASI Science Data Center (ASDC, Italy) and the California Institute of Technology (USA).

This research has made use of the XRT Data Analysis Software (XRTDAS). This work made use of data supplied by the UK Swift Science Data Centre at the University of Leicester. 

This research has made use of data obtained through the High Energy Astrophysics Science Archive Research Center Online Service, provided by the NASA/Goddard Space Flight Center.  This research has made use of the NASA/IPAC Extragalactic Database (NED), which is operated by the Jet Propulsion Laboratory, California Institute of Technology, under contract with the National Aeronautics and Space Administration. Part of this work is based on archival data, software or online services provided by the ASI Data Center (ASDC).
 
Data from the Steward Observatory spectropolarimetric monitoring project were used. This program is supported by Fermi Guest Investigator grants NNX08AW56G, NNX09AU10G, NNX12AO93G, and NNX15AU81G.
 
\bibliographystyle{aasjournal}
\bibliography{Master}

\begin{thebibliography}{}
\expandafter\ifx\csname natexlab\endcsname\relax\def\natexlab#1{#1}\fi

\bibitem[{{Abdo} {et~al.}(2010){Abdo}, {Ackermann}, {Ajello}, {Axelsson},
  {Baldini}, {Ballet}, {Barbiellini}, {Bastieri}, {Baughman}, {Bechtol}, \&
  et~al.}]{2010Natur.463..919A}
{Abdo}, A.~A., {Ackermann}, M., {Ajello}, M., {et~al.} 2010, \nat, 463, 919

\bibitem[{{Abdo} {et~al.}(2011){Abdo}, {Ackermann}, {Ajello}, {Allafort},
  {Baldini}, {Ballet}, {Barbiellini}, {Bastieri}, {Bellazzini}, {Berenji},
  {Blandford}, {Bloom}, {Bonamente}, {Borgland}, {Bouvier}, {Bregeon},
  {Brigida}, {Bruel}, {Buehler}, {Buson}, {Caliandro}, {Cameron}, {Caraveo},
  {Casandjian}, {Cavazzuti}, {Cecchi}, {Charles}, {Chekhtman}, {Cheung},
  {Chiang}, {Ciprini}, {Claus}, {Conrad}, {Cutini}, {D'Ammando}, {de Angelis},
  {de Palma}, {Dermer}, {Digel}, {Silva}, {Drell}, {Dubois}, {Dumora},
  {Escande}, {Favuzzi}, {Fegan}, {Ferrara}, {Fortin}, {Fukazawa}, {Fusco},
  {Gargano}, {Gasparrini}, {Gehrels}, {Germani}, {Giglietto}, {Giommi},
  {Giordano}, {Giroletti}, {Glanzman}, {Godfrey}, {Grenier}, {Grove},
  {Guiriec}, {Hadasch}, {Hayashida}, {Hays}, {Horan}, {Itoh},
  {J{\'o}hannesson}, {Johnson}, {Kamae}, {Katagiri}, {Kataoka},
  {Kn{\"o}dlseder}, {Kuss}, {Lande}, {Larsson}, {Latronico}, {Lee}, {Longo},
  {Loparco}, {Lott}, {Lovellette}, {Lubrano}, {Madejski}, {Makeev},
  {Mazziotta}, {McConville}, {McEnery}, {Michelson}, {Mitthumsiri}, {Mizuno},
  {Moiseev}, {Monte}, {Monzani}, {Morselli}, {Moskalenko}, {Murgia},
  {Naumann-Godo}, {Nishino}, {Nolan}, {Norris}, {Nuss}, {Ohsugi}, {Okumura},
  {Orlando}, {Ormes}, {Paneque}, {Pelassa}, {Pesce-Rollins}, {Pierbattista},
  {Piron}, {Porter}, {Rain{\`o}}, {Rando}, {Razzaque}, {Reimer}, {Reimer},
  {Ritz}, {Roth}, {Sadrozinski}, {Sanchez}, {Scargle}, {Schalk}, {Sgr{\`o}},
  {Siskind}, {Smith}, {Spandre}, {Spinelli}, {Strickman}, {Takahashi},
  {Takahashi}, {Tanaka}, {Tanaka}, {Thayer}, {Thayer}, {Thompson}, {Tibaldo},
  {Torres}, {Tosti}, {Tramacere}, {Troja}, {Vandenbroucke}, {Vasileiou},
  {Vianello}, {Vilchez}, {Vitale}, {Waite}, {Wang}, {Winer}, {Wood}, {Yang}, \&
  {Ziegler}}]{2011ApJ...733L..26A}
---. 2011, \apjl, 733, L26

\bibitem[{{Acero} {et~al.}(2015){Acero}, {Ackermann}, {Ajello}, {Albert},
  {Atwood}, {Axelsson}, {Baldini}, {Ballet}, {Barbiellini}, {Bastieri},
  {Belfiore}, {Bellazzini}, {Bissaldi}, {Blandford}, {Bloom}, {Bogart},
  {Bonino}, {Bottacini}, {Bregeon}, {Britto}, {Bruel}, {Buehler}, {Burnett},
  {Buson}, {Caliandro}, {Cameron}, {Caputo}, {Caragiulo}, {Caraveo},
  {Casandjian}, {Cavazzuti}, {Charles}, {Chaves}, {Chekhtman}, {Cheung},
  {Chiang}, {Chiaro}, {Ciprini}, {Claus}, {Cohen-Tanugi}, {Cominsky}, {Conrad},
  {Cutini}, {D'Ammando}, {de Angelis}, {DeKlotz}, {de Palma}, {Desiante},
  {Digel}, {Di Venere}, {Drell}, {Dubois}, {Dumora}, {Favuzzi}, {Fegan},
  {Ferrara}, {Finke}, {Franckowiak}, {Fukazawa}, {Funk}, {Fusco}, {Gargano},
  {Gasparrini}, {Giebels}, {Giglietto}, {Giommi}, {Giordano}, {Giroletti},
  {Glanzman}, {Godfrey}, {Grenier}, {Grondin}, {Grove}, {Guillemot}, {Guiriec},
  {Hadasch}, {Harding}, {Hays}, {Hewitt}, {Hill}, {Horan}, {Iafrate}, {Jogler},
  {J{\'o}hannesson}, {Johnson}, {Johnson}, {Johnson}, {Johnson}, {Kamae},
  {Kataoka}, {Katsuta}, {Kuss}, {La Mura}, {Landriu}, {Larsson}, {Latronico},
  {Lemoine-Goumard}, {Li}, {Li}, {Longo}, {Loparco}, {Lott}, {Lovellette},
  {Lubrano}, {Madejski}, {Massaro}, {Mayer}, {Mazziotta}, {McEnery},
  {Michelson}, {Mirabal}, {Mizuno}, {Moiseev}, {Mongelli}, {Monzani},
  {Morselli}, {Moskalenko}, {Murgia}, {Nuss}, {Ohno}, {Ohsugi}, {Omodei},
  {Orienti}, {Orlando}, {Ormes}, {Paneque}, {Panetta}, {Perkins},
  {Pesce-Rollins}, {Piron}, {Pivato}, {Porter}, {Racusin}, {Rando}, {Razzano},
  {Razzaque}, {Reimer}, {Reimer}, {Reposeur}, {Rochester}, {Romani},
  {Salvetti}, {S{\'a}nchez-Conde}, {Saz Parkinson}, {Schulz}, {Siskind},
  {Smith}, {Spada}, {Spandre}, {Spinelli}, {Stephens}, {Strong}, {Suson},
  {Takahashi}, {Takahashi}, {Tanaka}, {Thayer}, {Thayer}, {Thompson},
  {Tibaldo}, {Tibolla}, {Torres}, {Torresi}, {Tosti}, {Troja}, {Van Klaveren},
  {Vianello}, {Winer}, {Wood}, {Wood}, {Zimmer}, \& {Fermi-LAT
  Collaboration}}]{2015ApJS..218...23A}
{Acero}, F., {Ackermann}, M., {Ajello}, M., {et~al.} 2015, \apjs, 218, 23

\bibitem[{{Acero} {et~al.}(2016){Acero}, {Ackermann}, {Ajello}, {Albert},
  {Baldini}, {Ballet}, {Barbiellini}, {Bastieri}, {Bellazzini}, {Bissaldi},
  {Bloom}, {Bonino}, {Bottacini}, {Brandt}, {Bregeon}, {Bruel}, {Buehler},
  {Buson}, {Caliandro}, {Cameron}, {Caragiulo}, {Caraveo}, {Casandjian},
  {Cavazzuti}, {Cecchi}, {Charles}, {Chekhtman}, {Chiang}, {Chiaro}, {Ciprini},
  {Claus}, {Cohen-Tanugi}, {Conrad}, {Cuoco}, {Cutini}, {D'Ammando}, {de
  Angelis}, {de Palma}, {Desiante}, {Digel}, {Di Venere}, {Drell}, {Favuzzi},
  {Fegan}, {Ferrara}, {Focke}, {Franckowiak}, {Funk}, {Fusco}, {Gargano},
  {Gasparrini}, {Giglietto}, {Giordano}, {Giroletti}, {Glanzman}, {Godfrey},
  {Grenier}, {Guiriec}, {Hadasch}, {Harding}, {Hayashi}, {Hays}, {Hewitt},
  {Hill}, {Horan}, {Hou}, {Jogler}, {J{\'o}hannesson}, {Kamae}, {Kuss},
  {Landriu}, {Larsson}, {Latronico}, {Li}, {Li}, {Longo}, {Loparco},
  {Lovellette}, {Lubrano}, {Maldera}, {Malyshev}, {Manfreda}, {Martin},
  {Mayer}, {Mazziotta}, {McEnery}, {Michelson}, {Mirabal}, {Mizuno}, {Monzani},
  {Morselli}, {Nuss}, {Ohsugi}, {Omodei}, {Orienti}, {Orlando}, {Ormes},
  {Paneque}, {Pesce-Rollins}, {Piron}, {Pivato}, {Rain{\`o}}, {Rando},
  {Razzano}, {Razzaque}, {Reimer}, {Reimer}, {Remy}, {Renault},
  {S{\'a}nchez-Conde}, {Schaal}, {Schulz}, {Sgr{\`o}}, {Siskind}, {Spada},
  {Spandre}, {Spinelli}, {Strong}, {Suson}, {Tajima}, {Takahashi}, {Thayer},
  {Thompson}, {Tibaldo}, {Tinivella}, {Torres}, {Tosti}, {Troja}, {Vianello},
  {Werner}, {Wood}, {Wood}, {Zaharijas}, \& {Zimmer}}]{2016ApJS..223...26A}
---. 2016, \apjs, 223, 26

\bibitem[{{Ackermann} {et~al.}(2015){Ackermann}, {Ajello}, {Atwood}, {Baldini},
  {Ballet}, {Barbiellini}, {Bastieri}, {Becerra Gonzalez}, {Bellazzini},
  {Bissaldi}, {Blandford}, {Bloom}, {Bonino}, {Bottacini}, {Brandt}, {Bregeon},
  {Britto}, {Bruel}, {Buehler}, {Buson}, {Caliandro}, {Cameron}, {Caragiulo},
  {Caraveo}, {Carpenter}, {Casandjian}, {Cavazzuti}, {Cecchi}, {Charles},
  {Chekhtman}, {Cheung}, {Chiang}, {Chiaro}, {Ciprini}, {Claus},
  {Cohen-Tanugi}, {Cominsky}, {Conrad}, {Cutini}, {D'Abrusco}, {D'Ammando}, {de
  Angelis}, {Desiante}, {Digel}, {Di Venere}, {Drell}, {Favuzzi}, {Fegan},
  {Ferrara}, {Finke}, {Focke}, {Franckowiak}, {Fuhrmann}, {Fukazawa},
  {Furniss}, {Fusco}, {Gargano}, {Gasparrini}, {Giglietto}, {Giommi},
  {Giordano}, {Giroletti}, {Glanzman}, {Godfrey}, {Grenier}, {Grove},
  {Guiriec}, {Hewitt}, {Hill}, {Horan}, {Itoh}, {J{\'o}hannesson}, {Johnson},
  {Johnson}, {Kataoka}, {Kawano}, {Krauss}, {Kuss}, {La Mura}, {Larsson},
  {Latronico}, {Leto}, {Li}, {Li}, {Longo}, {Loparco}, {Lott}, {Lovellette},
  {Lubrano}, {Madejski}, {Mayer}, {Mazziotta}, {McEnery}, {Michelson},
  {Mizuno}, {Moiseev}, {Monzani}, {Morselli}, {Moskalenko}, {Murgia}, {Nuss},
  {Ohno}, {Ohsugi}, {Ojha}, {Omodei}, {Orienti}, {Orlando}, {Paggi}, {Paneque},
  {Perkins}, {Pesce-Rollins}, {Piron}, {Pivato}, {Porter}, {Rain{\`o}},
  {Rando}, {Razzano}, {Razzaque}, {Reimer}, {Reimer}, {Romani}, {Salvetti},
  {Schaal}, {Schinzel}, {Schulz}, {Sgr{\`o}}, {Siskind}, {Sokolovsky}, {Spada},
  {Spandre}, {Spinelli}, {Stawarz}, {Suson}, {Takahashi}, {Takahashi},
  {Tanaka}, {Thayer}, {Thayer}, {Tibaldo}, {Torres}, {Torresi}, {Tosti},
  {Troja}, {Uchiyama}, {Vianello}, {Winer}, {Wood}, \&
  {Zimmer}}]{2015ApJ...810...14A}
{Ackermann}, M., {Ajello}, M., {Atwood}, W.~B., {et~al.} 2015, \apj, 810, 14

\bibitem[{{Ackermann} {et~al.}(2017){Ackermann}, {Ajello}, {Baldini}, {Ballet},
  {Barbiellini}, {Bastieri}, {Becerra Gonzalez}, {Bellazzini}, {Bissaldi},
  {Blandford}, {Bloom}, {Bonino}, {Bottacini}, {Bregeon}, {Bruel}, {Buehler},
  {Buson}, {Cameron}, {Caragiulo}, {Caraveo}, {Cavazzuti}, {Cecchi}, {Cheung},
  {Chiang}, {Chiaro}, {Ciprini}, {Conrad}, {Costantin}, {Costanza}, {Cutini},
  {D'Ammando}, {de Palma}, {Desiante}, {Digel}, {Di Lalla}, {Di Mauro}, {Di
  Venere}, {Dom{\'{\i}}nguez}, {Drell}, {Favuzzi}, {Fegan}, {Ferrara}, {Finke},
  {Focke}, {Fukazawa}, {Funk}, {Fusco}, {Gargano}, {Gasparrini}, {Giglietto},
  {Giordano}, {Giroletti}, {Green}, {Grenier}, {Guillemot}, {Guiriec},
  {Hartmann}, {Hays}, {Horan}, {Jogler}, {J{\'o}hannesson}, {Johnson}, {Kuss},
  {La Mura}, {Larsson}, {Latronico}, {Li}, {Longo}, {Loparco}, {Lovellette},
  {Lubrano}, {Magill}, {Maldera}, {Manfreda}, {Marcotulli}, {Mazziotta},
  {Michelson}, {Mirabal}, {Mitthumsiri}, {Mizuno}, {Monzani}, {Morselli},
  {Moskalenko}, {Negro}, {Nuss}, {Ohsugi}, {Ojha}, {Omodei}, {Orienti},
  {Orlando}, {Ormes}, {Paliya}, {Paneque}, {Perkins}, {Persic},
  {Pesce-Rollins}, {Piron}, {Porter}, {Principe}, {Rain{\`o}}, {Rando}, {Rani},
  {Razzano}, {Razzaque}, {Reimer}, {Reimer}, {Romani}, {Sgr{\`o}}, {Simone},
  {Siskind}, {Spada}, {Spandre}, {Spinelli}, {Stalin}, {Stawarz}, {Suson},
  {Takahashi}, {Tanaka}, {Thayer}, {Thompson}, {Torres}, {Torresi}, {Tosti},
  {Troja}, {Vianello}, \& {Wood}}]{2017ApJ...837L...5A}
{Ackermann}, M., {Ajello}, M., {Baldini}, L., {et~al.} 2017, \apjl, 837, L5

\bibitem[{{Ajello} {et~al.}(2009){Ajello}, {Costamante}, {Sambruna}, {Gehrels},
  {Chiang}, {Rau}, {Escala}, {Greiner}, {Tueller}, {Wall}, \&
  {Mushotzky}}]{2009ApJ...699..603A}
{Ajello}, M., {Costamante}, L., {Sambruna}, R.~M., {et~al.} 2009, \apj, 699,
  603

\bibitem[{{Ajello} {et~al.}(2016){Ajello}, {Ghisellini}, {Paliya}, {Kocevski},
  {Tagliaferri}, {Madejski}, {Rau}, {Schady}, {Greiner}, {Massaro},
  {Balokovi{\'c}}, {B{\"u}hler}, {Giomi}, {Marcotulli}, {D'Ammando}, {Stern},
  {Boggs}, {Christensen}, {Craig}, {Hailey}, {Harrison}, \&
  {Zhang}}]{2016ApJ...826...76A}
{Ajello}, M., {Ghisellini}, G., {Paliya}, V.~S., {et~al.} 2016, \apj, 826, 76

\bibitem[{{Angioni} \& {Cheung}(2018)}]{2018ATel11137....1A}
{Angioni}, R., \& {Cheung}, C.~C. 2018, The Astronomer's Telegram, 11137, 1

\bibitem[{{Arcodia} {et~al.}(2018){Arcodia}, {Campana}, {Salvaterra}, \&
  {Ghisellini}}]{2018A&A...616A.170A}
{Arcodia}, R., {Campana}, S., {Salvaterra}, R., \& {Ghisellini}, G. 2018, \aap,
  616, A170

\bibitem[{{Arnaud}(1996)}]{1996ASPC..101...17A}
{Arnaud}, K.~A. 1996, in Astronomical Society of the Pacific Conference Series,
  Vol. 101, Astronomical Data Analysis Software and Systems V, ed. G.~H.
  {Jacoby} \& J.~{Barnes}, 17

\bibitem[{{Atwood} {et~al.}(2013){Atwood}, {Albert}, {Baldini}, {Tinivella},
  {Bregeon}, {Pesce-Rollins}, {Sgr{\`o}}, {Bruel}, {Charles}, {Drlica-Wagner},
  {Franckowiak}, {Jogler}, {Rochester}, {Usher}, {Wood}, {Cohen-Tanugi}, \&
  {S.~Zimmer for the Fermi-LAT Collaboration}}]{2013arXiv1303.3514A}
{Atwood}, W., {Albert}, A., {Baldini}, L., {et~al.} 2013, ArXiv e-prints,
  arXiv:1303.3514

\bibitem[{{Atwood} {et~al.}(2009){Atwood}, {Abdo}, {Ackermann}, {Althouse},
  {Anderson}, {Axelsson}, {Baldini}, {Ballet}, {Band}, {Barbiellini}, \&
  et~al.}]{2009ApJ...697.1071A}
{Atwood}, W.~B., {Abdo}, A.~A., {Ackermann}, M., {et~al.} 2009, \apj, 697, 1071

\bibitem[{{Barthelmy} {et~al.}(2005){Barthelmy}, {Barbier}, {Cummings},
  {Fenimore}, {Gehrels}, {Hullinger}, {Krimm}, {Markwardt}, {Palmer},
  {Parsons}, {Sato}, {Suzuki}, {Takahashi}, {Tashiro}, \&
  {Tueller}}]{2005SSRv..120..143B}
{Barthelmy}, S.~D., {Barbier}, L.~M., {Cummings}, J.~R., {et~al.} 2005, \ssr,
  120, 143

\bibitem[{{Bloemen} {et~al.}(1995){Bloemen}, {Bennett}, {Blom}, {Collmar},
  {Hermsen}, {Lichti}, {Morris}, {Schoenfelder}, {Stacy}, {Strong}, \&
  {Winkler}}]{1995A&A...293L...1B}
{Bloemen}, H., {Bennett}, K., {Blom}, J.~J., {et~al.} 1995, \aap, 293

\bibitem[{{Breeveld} {et~al.}(2011){Breeveld}, {Landsman}, {Holland}, {Roming},
  {Kuin}, \& {Page}}]{2011AIPC.1358..373B}
{Breeveld}, A.~A., {Landsman}, W., {Holland}, S.~T., {et~al.} 2011, in American
  Institute of Physics Conference Series, Vol. 1358, American Institute of
  Physics Conference Series, ed. J.~E. {McEnery}, J.~L. {Racusin}, \&
  N.~{Gehrels}, 373--376

\bibitem[{{Burrows} {et~al.}(2005){Burrows}, {Hill}, {Nousek}, {Kennea},
  {Wells}, {Osborne}, {Abbey}, {Beardmore}, {Mukerjee}, {Short}, {Chincarini},
  {Campana}, {Citterio}, {Moretti}, {Pagani}, {Tagliaferri}, {Giommi},
  {Capalbi}, {Tamburelli}, {Angelini}, {Cusumano}, {Br{\"a}uninger}, {Burkert},
  \& {Hartner}}]{2005SSRv..120..165B}
{Burrows}, D.~N., {Hill}, J.~E., {Nousek}, J.~A., {et~al.} 2005, \ssr, 120, 165

\bibitem[{{Celotti} \& {Ghisellini}(2008)}]{2008MNRAS.385..283C}
{Celotti}, A., \& {Ghisellini}, G. 2008, \mnras, 385, 283

\bibitem[{{Celotti} {et~al.}(1997){Celotti}, {Padovani}, \&
  {Ghisellini}}]{1997MNRAS.286..415C}
{Celotti}, A., {Padovani}, P., \& {Ghisellini}, G. 1997, \mnras, 286, 415

\bibitem[{{Condon} {et~al.}(1998){Condon}, {Cotton}, {Greisen}, {Yin},
  {Perley}, {Taylor}, \& {Broderick}}]{1998AJ....115.1693C}
{Condon}, J.~J., {Cotton}, W.~D., {Greisen}, E.~W., {et~al.} 1998, \aj, 115,
  1693

\bibitem[{{Costamante} {et~al.}(2018){Costamante}, {Cutini}, {Tosti},
  {Antolini}, \& {Tramacere}}]{2018MNRAS.477.4749C}
{Costamante}, L., {Cutini}, S., {Tosti}, G., {Antolini}, E., \& {Tramacere}, A.
  2018, \mnras, 477, 4749

\bibitem[{{Dermer} {et~al.}(2009){Dermer}, {Finke}, {Krug}, \&
  {B{\"o}ttcher}}]{2009ApJ...692...32D}
{Dermer}, C.~D., {Finke}, J.~D., {Krug}, H., \& {B{\"o}ttcher}, M. 2009, \apj,
  692, 32

\bibitem[{{Donato} {et~al.}(2001){Donato}, {Ghisellini}, {Tagliaferri}, \&
  {Fossati}}]{2001A&A...375..739D}
{Donato}, D., {Ghisellini}, G., {Tagliaferri}, G., \& {Fossati}, G. 2001, \aap,
  375, 739

\bibitem[{{Eitan} \& {Behar}(2013)}]{2013ApJ...774...29E}
{Eitan}, A., \& {Behar}, E. 2013, \apj, 774, 29

\bibitem[{{Fabian} {et~al.}(2001){Fabian}, {Celotti}, {Iwasawa}, {McMahon},
  {Carilli}, {Brandt}, {Ghisellini}, \& {Hook}}]{2001MNRAS.323..373F}
{Fabian}, A.~C., {Celotti}, A., {Iwasawa}, K., {et~al.} 2001, \mnras, 323, 373

\bibitem[{{Finke} {et~al.}(2008){Finke}, {Dermer}, \&
  {B{\"o}ttcher}}]{2008ApJ...686..181F}
{Finke}, J.~D., {Dermer}, C.~D., \& {B{\"o}ttcher}, M. 2008, \apj, 686, 181

\bibitem[{{Finke} {et~al.}(2010){Finke}, {Razzaque}, \&
  {Dermer}}]{2010ApJ...712..238F}
{Finke}, J.~D., {Razzaque}, S., \& {Dermer}, C.~D. 2010, \apj, 712, 238

\bibitem[{{Fossati} {et~al.}(1998){Fossati}, {Maraschi}, {Celotti}, {Comastri},
  \& {Ghisellini}}]{1998MNRAS.299..433F}
{Fossati}, G., {Maraschi}, L., {Celotti}, A., {Comastri}, A., \& {Ghisellini},
  G. 1998, \mnras, 299, 433

\bibitem[{{Francis} {et~al.}(1991){Francis}, {Hewett}, {Foltz}, {Chaffee},
  {Weymann}, \& {Morris}}]{1991ApJ...373..465F}
{Francis}, P.~J., {Hewett}, P.~C., {Foltz}, C.~B., {et~al.} 1991, \apj, 373,
  465

\bibitem[{{Frank} {et~al.}(2002){Frank}, {King}, \&
  {Raine}}]{2002apa..book.....F}
{Frank}, J., {King}, A., \& {Raine}, D.~J. 2002, {Accretion Power in
  Astrophysics, by Juhan Frank and Andrew King and Derek Raine, pp.~398.~ISBN
  0521620538.~Cambridge, UK: Cambridge University Press, February 2002}

\bibitem[{{Ghisellini} {et~al.}(1998){Ghisellini}, {Celotti}, {Fossati},
  {Maraschi}, \& {Comastri}}]{1998MNRAS.301..451G}
{Ghisellini}, G., {Celotti}, A., {Fossati}, G., {Maraschi}, L., \& {Comastri},
  A. 1998, \mnras, 301, 451

\bibitem[{{Ghisellini} \& {Tavecchio}(2009)}]{2009MNRAS.397..985G}
{Ghisellini}, G., \& {Tavecchio}, F. 2009, \mnras, 397, 985

\bibitem[{{Ghisellini} {et~al.}(2011){Ghisellini}, {Tavecchio}, {Foschini}, \&
  {Ghirlanda}}]{2011MNRAS.414.2674G}
{Ghisellini}, G., {Tavecchio}, F., {Foschini}, L., \& {Ghirlanda}, G. 2011,
  \mnras, 414, 2674

\bibitem[{{Ghisellini} {et~al.}(2010){Ghisellini}, {Della Ceca}, {Volonteri},
  {Ghirlanda}, {Tavecchio}, {Foschini}, {Tagliaferri}, {Haardt}, {Pareschi}, \&
  {Grindlay}}]{2010MNRAS.405..387G}
{Ghisellini}, G., {Della Ceca}, R., {Volonteri}, M., {et~al.} 2010, \mnras,
  405, 387

\bibitem[{{Giommi} {et~al.}(2012){Giommi}, {Padovani}, {Polenta}, {Turriziani},
  {D'Elia}, \& {Piranomonte}}]{2012MNRAS.420.2899G}
{Giommi}, P., {Padovani}, P., {Polenta}, G., {et~al.} 2012, \mnras, 420, 2899

\bibitem[{{Harrison} {et~al.}(2013){Harrison}, {Craig}, {Christensen},
  {Hailey}, {Zhang}, {Boggs}, {Stern}, {Cook}, {Forster}, {Giommi},
  {Grefenstette}, {Kim}, {Kitaguchi}, {Koglin}, {Madsen}, {Mao}, {Miyasaka},
  {Mori}, {Perri}, {Pivovaroff}, {Puccetti}, {Rana}, {Westergaard}, {Willis},
  {Zoglauer}, {An}, {Bachetti}, {Barri{\`e}re}, {Bellm}, {Bhalerao},
  {Brejnholt}, {Fuerst}, {Liebe}, {Markwardt}, {Nynka}, {Vogel}, {Walton},
  {Wik}, {Alexander}, {Cominsky}, {Hornschemeier}, {Hornstrup}, {Kaspi},
  {Madejski}, {Matt}, {Molendi}, {Smith}, {Tomsick}, {Ajello}, {Ballantyne},
  {Balokovi{\'c}}, {Barret}, {Bauer}, {Blandford}, {Brandt}, {Brenneman},
  {Chiang}, {Chakrabarty}, {Chenevez}, {Comastri}, {Dufour}, {Elvis}, {Fabian},
  {Farrah}, {Fryer}, {Gotthelf}, {Grindlay}, {Helfand}, {Krivonos}, {Meier},
  {Miller}, {Natalucci}, {Ogle}, {Ofek}, {Ptak}, {Reynolds}, {Rigby},
  {Tagliaferri}, {Thorsett}, {Treister}, \& {Urry}}]{2013ApJ...770..103H}
{Harrison}, F.~A., {Craig}, W.~W., {Christensen}, F.~E., {et~al.} 2013, \apj,
  770, 103

\bibitem[{{Healey} {et~al.}(2008){Healey}, {Romani}, {Cotter}, {Michelson},
  {Schlafly}, {Readhead}, {Giommi}, {Chaty}, {Grenier}, \&
  {Weintraub}}]{2008ApJS..175...97H}
{Healey}, S.~E., {Romani}, R.~W., {Cotter}, G., {et~al.} 2008, \apjs, 175, 97

\bibitem[{{Hogg} {et~al.}(2002){Hogg}, {Baldry}, {Blanton}, \&
  {Eisenstein}}]{2002astro.ph.10394H}
{Hogg}, D.~W., {Baldry}, I.~K., {Blanton}, M.~R., \& {Eisenstein}, D.~J. 2002,
  ArXiv Astrophysics e-prints, astro-ph/0210394

\bibitem[{{Jorstad} {et~al.}(2005){Jorstad}, {Marscher}, {Lister}, {Stirling},
  {Cawthorne}, {Gear}, {G{\'o}mez}, {Stevens}, {Smith}, {Forster}, \&
  {Robson}}]{2005AJ....130.1418J}
{Jorstad}, S.~G., {Marscher}, A.~P., {Lister}, M.~L., {et~al.} 2005, \aj, 130,
  1418

\bibitem[{{Kalberla} {et~al.}(2005){Kalberla}, {Burton}, {Hartmann}, {Arnal},
  {Bajaja}, {Morras}, \& {P{\"o}ppel}}]{2005AA...440..775K}
{Kalberla}, P.~M.~W., {Burton}, W.~B., {Hartmann}, D., {et~al.} 2005, \aap,
  440, 775

\bibitem[{{Kaur} {et~al.}(2018){Kaur}, {Rau}, {Ajello}, {Dom{\'{\i}}nguez},
  {Paliya}, {Greiner}, {Hartmann}, \& {Schady}}]{2018ApJ...859...80K}
{Kaur}, A., {Rau}, A., {Ajello}, M., {et~al.} 2018, \apj, 859, 80

\bibitem[{{Kaur} {et~al.}(2017){Kaur}, {Rau}, {Ajello}, {Greiner}, {Hartmann},
  {Paliya}, {Dom{\'{\i}}nguez}, {Bolmer}, \& {Schady}}]{2017ApJ...834...41K}
---. 2017, \apj, 834, 41

\bibitem[{{Larionov} {et~al.}(2013){Larionov}, {Jorstad}, {Marscher},
  {Morozova}, {Blinov}, {Hagen-Thorn}, {Konstantinova}, {Kopatskaya},
  {Larionova}, {Larionova}, \& {Troitsky}}]{2013ApJ...768...40L}
{Larionov}, V.~M., {Jorstad}, S.~G., {Marscher}, A.~P., {et~al.} 2013, \apj,
  768, 40

\bibitem[{{Lei} \& {Wang}(2014)}]{2014PASJ...66...92L}
{Lei}, M., \& {Wang}, J. 2014, \pasj, 66, 92

\bibitem[{{Lister} {et~al.}(2016){Lister}, {Aller}, {Aller}, {Homan},
  {Kellermann}, {Kovalev}, {Pushkarev}, {Richards}, {Ros}, \&
  {Savolainen}}]{2016AJ....152...12L}
{Lister}, M.~L., {Aller}, M.~F., {Aller}, H.~D., {et~al.} 2016, \aj, 152, 12

\bibitem[{{Marcotulli} {et~al.}(2017){Marcotulli}, {Paliya}, {Ajello}, {Kaur},
  {Hartmann}, {Gasparrini}, {Greiner}, {Rau}, {Schady}, {Balokovi{\'c}},
  {Stern}, \& {Madejski}}]{2017ApJ...839...96M}
{Marcotulli}, L., {Paliya}, V.~S., {Ajello}, M., {et~al.} 2017, \apj, 839, 96

\bibitem[{{Marscher} {et~al.}(2008){Marscher}, {Jorstad}, {D'Arcangelo},
  {Smith}, {Williams}, {Larionov}, {Oh}, {Olmstead}, {Aller}, {Aller},
  {McHardy}, {L{\"a}hteenm{\"a}ki}, {Tornikoski}, {Valtaoja}, {Hagen-Thorn},
  {Kopatskaya}, {Gear}, {Tosti}, {Kurtanidze}, {Nikolashvili}, {Sigua},
  {Miller}, \& {Ryle}}]{2008Natur.452..966M}
{Marscher}, A.~P., {Jorstad}, S.~G., {D'Arcangelo}, F.~D., {et~al.} 2008, \nat,
  452, 966

\bibitem[{{Mattox} {et~al.}(1996){Mattox}, {Bertsch}, {Chiang}, {Dingus},
  {Digel}, {Esposito}, {Fierro}, {Hartman}, {Hunter}, {Kanbach}, {Kniffen},
  {Lin}, {Macomb}, {Mayer-Hasselwander}, {Michelson}, {von Montigny},
  {Mukherjee}, {Nolan}, {Ramanamurthy}, {Schneid}, {Sreekumar}, {Thompson}, \&
  {Willis}}]{1996ApJ...461..396M}
{Mattox}, J.~R., {Bertsch}, D.~L., {Chiang}, J., {et~al.} 1996, \apj, 461, 396

\bibitem[{{McIntosh} {et~al.}(1999){McIntosh}, {Rieke}, {Rix}, {Foltz}, \&
  {Weymann}}]{1999ApJ...514...40M}
{McIntosh}, D.~H., {Rieke}, M.~J., {Rix}, H.-W., {Foltz}, C.~B., \& {Weymann},
  R.~J. 1999, \apj, 514, 40

\bibitem[{{Nolan} {et~al.}(2012){Nolan}, {Abdo}, {Ackermann}, {Ajello},
  {Allafort}, {Antolini}, {Atwood}, {Axelsson}, {Baldini}, {Ballet}, \&
  et~al.}]{2012ApJS..199...31N}
{Nolan}, P.~L., {Abdo}, A.~A., {Ackermann}, M., {et~al.} 2012, \apjs, 199, 31

\bibitem[{{O'Dea} {et~al.}(1990){O'Dea}, {Baum}, {Stanghellini}, {Morris},
  {Patnaik}, \& {Gopal-Krishna}}]{1990A&AS...84..549O}
{O'Dea}, C.~P., {Baum}, S.~A., {Stanghellini}, C., {et~al.} 1990, \aaps, 84,
  549

\bibitem[{{Oh} {et~al.}(2018){Oh}, {Koss}, {Markwardt}, {Schawinski},
  {Baumgartner}, {Barthelmy}, {Cenko}, {Gehrels}, {Mushotzky}, {Petulante},
  {Ricci}, {Lien}, \& {Trakhtenbrot}}]{2018ApJS..235....4O}
{Oh}, K., {Koss}, M., {Markwardt}, C.~B., {et~al.} 2018, \apjs, 235, 4

\bibitem[{{Orienti} {et~al.}(2014){Orienti}, {D'Ammando}, {Giroletti}, {Finke},
  {Ajello}, {Dallacasa}, \& {Venturi}}]{2014MNRAS.444.3040O}
{Orienti}, M., {D'Ammando}, F., {Giroletti}, M., {et~al.} 2014, \mnras, 444,
  3040

\bibitem[{{Paliya}(2015)}]{2015ApJ...804...74P}
{Paliya}, V.~S. 2015, \apj, 804, 74

\bibitem[{{Paliya} {et~al.}(2017){Paliya}, {Marcotulli}, {Ajello}, {Joshi},
  {Sahayanathan}, {Rao}, \& {Hartmann}}]{2017ApJ...851...33P}
{Paliya}, V.~S., {Marcotulli}, L., {Ajello}, M., {et~al.} 2017, \apj, 851, 33

\bibitem[{{Paliya} {et~al.}(2016){Paliya}, {Parker}, {Fabian}, \&
  {Stalin}}]{2016ApJ...825...74P}
{Paliya}, V.~S., {Parker}, M.~L., {Fabian}, A.~C., \& {Stalin}, C.~S. 2016,
  \apj, 825, 74

\bibitem[{{Planck Collaboration} {et~al.}(2016){Planck Collaboration}, {Ade},
  {Aghanim}, {Arnaud}, {Ashdown}, {Aumont}, {Baccigalupi}, {Banday},
  {Barreiro}, {Bartlett}, \& et~al.}]{2016A&A...594A..13P}
{Planck Collaboration}, {Ade}, P.~A.~R., {Aghanim}, N., {et~al.} 2016, \aap,
  594, A13

\bibitem[{{Roming} {et~al.}(2005){Roming}, {Kennedy}, {Mason}, {Nousek}, {Ahr},
  {Bingham}, {Broos}, {Carter}, {Hancock}, {Huckle}, {Hunsberger}, {Kawakami},
  {Killough}, {Koch}, {McLelland}, {Smith}, {Smith}, {Soto}, {Boyd},
  {Breeveld}, {Holland}, {Ivanushkina}, {Pryzby}, {Still}, \&
  {Stock}}]{2005SSRv..120...95R}
{Roming}, P.~W.~A., {Kennedy}, T.~E., {Mason}, K.~O., {et~al.} 2005, \ssr, 120,
  95

\bibitem[{{Sbarrato} {et~al.}(2016){Sbarrato}, {Ghisellini}, {Tagliaferri},
  {Perri}, {Madejski}, {Stern}, {Boggs}, {Christensen}, {Craig}, {Hailey},
  {Harrison}, \& {Zhang}}]{2016MNRAS.462.1542S}
{Sbarrato}, T., {Ghisellini}, G., {Tagliaferri}, G., {et~al.} 2016, \mnras,
  462, 1542

\bibitem[{{Schilizzi} \& {Shaver}(1981)}]{1981A&A....96..365S}
{Schilizzi}, R.~T., \& {Shaver}, P.~A. 1981, \aap, 96, 365

\bibitem[{{Schlafly} \& {Finkbeiner}(2011)}]{2011ApJ...737..103S}
{Schlafly}, E.~F., \& {Finkbeiner}, D.~P. 2011, \apj, 737, 103

\bibitem[{{Shakura} \& {Sunyaev}(1973)}]{1973A&A....24..337S}
{Shakura}, N.~I., \& {Sunyaev}, R.~A. 1973, \aap, 24, 337

\bibitem[{{Shen} {et~al.}(2011){Shen}, {Richards}, {Strauss}, {Hall},
  {Schneider}, {Snedden}, {Bizyaev}, {Brewington}, {Malanushenko},
  {Malanushenko}, {Oravetz}, {Pan}, \& {Simmons}}]{2011ApJS..194...45S}
{Shen}, Y., {Richards}, G.~T., {Strauss}, M.~A., {et~al.} 2011, \apjs, 194, 45

\bibitem[{{Skrutskie} {et~al.}(2006){Skrutskie}, {Cutri}, {Stiening},
  {Weinberg}, {Schneider}, {Carpenter}, {Beichman}, {Capps}, {Chester},
  {Elias}, {Huchra}, {Liebert}, {Lonsdale}, {Monet}, {Price}, {Seitzer},
  {Jarrett}, {Kirkpatrick}, {Gizis}, {Howard}, {Evans}, {Fowler}, {Fullmer},
  {Hurt}, {Light}, {Kopan}, {Marsh}, {McCallon}, {Tam}, {Van Dyk}, \&
  {Wheelock}}]{2006AJ....131.1163S}
{Skrutskie}, M.~F., {Cutri}, R.~M., {Stiening}, R., {et~al.} 2006, \aj, 131,
  1163

\bibitem[{{Smith} {et~al.}(2009){Smith}, {Montiel}, {Rightley}, {Turner},
  {Schmidt}, \& {Jannuzi}}]{2009arXiv0912.3621S}
{Smith}, P.~S., {Montiel}, E., {Rightley}, S., {et~al.} 2009, arXiv:0912.3621,
  arXiv:0912.3621

\bibitem[{{Spangler} {et~al.}(1983){Spangler}, {Mutel}, \&
  {Benson}}]{1983ApJ...271...44S}
{Spangler}, S.~R., {Mutel}, R.~L., \& {Benson}, J.~M. 1983, \apj, 271, 44

\bibitem[{{Stickel} {et~al.}(1991){Stickel}, {Padovani}, {Urry}, {Fried}, \&
  {Kuehr}}]{1991ApJ...374..431S}
{Stickel}, M., {Padovani}, P., {Urry}, C.~M., {Fried}, J.~W., \& {Kuehr}, H.
  1991, \apj, 374, 431

\bibitem[{{Tagliaferri} {et~al.}(2015){Tagliaferri}, {Ghisellini}, {Perri},
  {Hayashida}, {Balokovi{\'c}}, {Covino}, {Giommi}, {Madejski}, {Puccetti},
  {Sbarrato}, {Boggs}, {Chiang}, {Christensen}, {Craig}, {Hailey}, {Harrison},
  {Stern}, \& {Zhang}}]{2015ApJ...807..167T}
{Tagliaferri}, G., {Ghisellini}, G., {Perri}, M., {et~al.} 2015, \apj, 807, 167

\bibitem[{{Tavecchio} {et~al.}(2010){Tavecchio}, {Ghisellini}, {Bonnoli}, \&
  {Ghirlanda}}]{2010MNRAS.405L..94T}
{Tavecchio}, F., {Ghisellini}, G., {Bonnoli}, G., \& {Ghirlanda}, G. 2010,
  \mnras, 405, L94

\bibitem[{{Torrealba} {et~al.}(2012){Torrealba}, {Chavushyan},
  {Cruz-Gonz{\'a}lez}, {Arshakian}, {Bertone}, \&
  {Rosa-Gonz{\'a}lez}}]{2012RMxAA..48....9T}
{Torrealba}, J., {Chavushyan}, V., {Cruz-Gonz{\'a}lez}, I., {et~al.} 2012,
  \rmxaa, 48, 9

\bibitem[{{Wang} {et~al.}(2001){Wang}, {Hong}, {Jiang}, {Venturi}, {Chen}, \&
  {An}}]{2001A&A...380..123W}
{Wang}, W.~H., {Hong}, X.~Y., {Jiang}, D.~R., {et~al.} 2001, \aap, 380, 123

\bibitem[{{Wehrle} {et~al.}(2012){Wehrle}, {Marscher}, {Jorstad}, {Gurwell},
  {Joshi}, {MacDonald}, {Williamson}, {Agudo}, \&
  {Grupe}}]{2012ApJ...758...72W}
{Wehrle}, A.~E., {Marscher}, A.~P., {Jorstad}, S.~G., {et~al.} 2012, \apj, 758,
  72

\bibitem[{{Whiting}(2005)}]{2005MmSAI..76...61W}
{Whiting}, M.~T. 2005, \memsai, 76, 61

\bibitem[{{Wills} {et~al.}(2011){Wills}, {Wills}, \&
  {Breger}}]{2011ApJS..194...19W}
{Wills}, B.~J., {Wills}, D., \& {Breger}, M. 2011, \apjs, 194, 19

\bibitem[{{Wills} \& {Wills}(1976)}]{1976ApJS...31..143W}
{Wills}, D., \& {Wills}, B.~J. 1976, \apjs, 31, 143

\bibitem[{{Worsley} {et~al.}(2004){Worsley}, {Fabian}, {Celotti}, \&
  {Iwasawa}}]{2004MNRAS.350L..67W}
{Worsley}, M.~A., {Fabian}, A.~C., {Celotti}, A., \& {Iwasawa}, K. 2004,
  \mnras, 350, L67

\bibitem[{{Wright} {et~al.}(2010){Wright}, {Eisenhardt}, {Mainzer}, {Ressler},
  {Cutri}, {Jarrett}, {Kirkpatrick}, {Padgett}, {McMillan}, {Skrutskie},
  {Stanford}, {Cohen}, {Walker}, {Mather}, {Leisawitz}, {Gautier}, {McLean},
  {Benford}, {Lonsdale}, {Blain}, {Mendez}, {Irace}, {Duval}, {Liu}, {Royer},
  {Heinrichsen}, {Howard}, {Shannon}, {Kendall}, {Walsh}, {Larsen}, {Cardon},
  {Schick}, {Schwalm}, {Abid}, {Fabinsky}, {Naes}, \&
  {Tsai}}]{2010AJ....140.1868W}
{Wright}, E.~L., {Eisenhardt}, P.~R.~M., {Mainzer}, A.~K., {et~al.} 2010, \aj,
  140, 1868

\bibitem[{{Yuan} \& {Wills}(2003)}]{2003ApJ...593L..11Y}
{Yuan}, M.~J., \& {Wills}, B.~J. 2003, \apjl, 593, L11

\bibitem[{{Zdziarski} {et~al.}(2015){Zdziarski}, {Sikora}, {Pjanka}, \&
  {Tchekhovskoy}}]{2015MNRAS.451..927Z}
{Zdziarski}, A.~A., {Sikora}, M., {Pjanka}, P., \& {Tchekhovskoy}, A. 2015,
  \mnras, 451, 927

\end{thebibliography}

\begin{table*}[h!]
\begin{center}
\caption{Summary of the SED analysis.\label{tab:sed_par}}
\begin{tabular}{rccccc}
\tableline\tableline
 & & \fermi-LAT & &\\
 Activity state & Time bin & $\Gamma_{0.1-300~{\rm GeV}}$ & $F_{0.1-300~{\rm GeV}}$                               & TS\\
                      & (MJD)  &                                                 & (10$^{-8}$ ph cm$^{-2}$ s$^{-1}$)  &                        \\ 
\tableline
 Flare  & 58122$-$58128 & 2.32$\pm$0.10 & 79.6$\pm$10.2 & 235\\
 Low activity    & 54683$-$58118 & 2.88$\pm$0.06 & 4.8$\pm$0.6 & 566\\
 \tableline
 & & \nustar~+ \swift-XRT & &\\
 Activity state &  & $\Gamma_{0.3-79~{\rm keV}}$ & Norm$_{0.3-79~{\rm keV}}$                               & Statistics\\
                      &  &                                                 & (10$^{-4}$ ph cm$^{-2}$ s$^{-1}$ keV$^{-1}$)  & $\chi^2$/dof \\ 
\tableline
 Flare &  & 1.58$^{+0.07}_{-0.08}$ & 4.93$^{+0.85}_{-0.73}$ & 150.73/120\\
 \tableline
 & & \nustar  & & \\
 Activity state & Exp.    & $\Gamma_{3-79~{\rm keV}}$ & Norm$_{3-79~{\rm keV}}$                              & Stat.\\
                      & (ksec) &                                                  & (10$^{-4}$ ph cm$^{-2}$ s$^{-1}$ keV$^{-1}$) & $\chi^2$/dof \\ 
 \tableline
 Flare & 28.56 & 1.57$^{+0.08}_{-0.07}$  & 4.89$^{+0.89}_{-0.72}$ & 141.92/109\\
\tableline
& & \swift-XRT  & & \\
 Activity state & Exp. & $\Gamma_{0.3-10~{\rm keV}}$ & Norm$_{0.3-10~{\rm keV}}$ & Stat.\\
                       & (ksec) &                                              & (10$^{-4}$ ph cm$^{-2}$ s$^{-1}$ keV$^{-1}$) & $\chi^2$/dof \\ 
\tableline
 Flare            & 5.64   & 1.62$^{+0.25}_{-0.24}$  & 4.62$^{+0.92}_{-0.87}$ & 8.72/10\\
 Low activity & 17.64 & 1.22$^{+0.12}_{-0.12}$  & 2.44$^{+0.30}_{-0.29}$ & 24.23/26\\
 \tableline
 & & \swift-UVOT &  & \\
 Activity state & $V$ & $B$ & $U$ & $UVW1$ \\
                      & (10$^{-12}$ \ergflux) & (10$^{-12}$ \ergflux) & (10$^{-12}$ \ergflux) & (10$^{-12}$ \ergflux) & \\
  \tableline
 Flare            & 4.96$\pm$0.18 & 5.71$\pm$0.34 & 4.04$\pm$0.21 & ---\\
 Low activity & 4.58$\pm$0.27 & 5.07$\pm$0.26 & 4.07$\pm$0.23 & 1.24$\pm$0.23\\
 \tableline
\end{tabular}
\end{center}
\tablecomments{For the low activity state, we combine a total of seven \swift~observations (observation id: 00036315001$-$2, 00036315004$-$8). On the other hand, flaring state X-ray spectrum is generated by adding three ToO observations (observation id: 00036315009$-$11).}
\end{table*}

\begin{table*}[t!]
{\small
\begin{center}
\caption{Summary of the parameters used/derived from the modeling of the SEDs. The adopted time-periods are defined in Table \ref{tab:sed_par}. We consider the central black hole mass and the accretion disk lumiunosity of $3\times10^9$ \msun~and $3\times10^{47}$ \lum, respectively (Section \ref{subsec:bh}) and assume the characteristic temperature of the IR-torus as 400 K. A viewing angle of 2$^{\circ}$ is adopted which is consistent with that typically observed from blazars using radio studies \citep[][]{2005AJ....130.1418J,2016AJ....152...12L}. Note that the jet powers are computed by assuming a two-sided jet.}\label{tab:sed}
\begin{tabular}{lccc}
\tableline
\tableline
Parameter                                            &  Symbol                     &    Low Activity (Q)     &   High activity (F) \\
\tableline
Slope of particle spectral index before break energy & {\it p}                          & 1.7        & 1.8      \\
Slope of particle spectral index after break energy   & {\it q}                          & 4.5         & 3.7      \\
Minimum Lorentz factor of the particle distribution  & $\gamma'_{\rm min}$   & 1           & 1          \\
Break Lorentz factor of the particle distribution      & $\gamma'_{\rm bk}$     & 145       & 323    \\
Maximum Lorentz factor of the particle distribution  & $\gamma'_{\rm max}$ & 3000     & 3500   \\
Magnetic field, in Gauss                                          & {$B$}                           & 0.6        & 0.6     \\
Bulk Lorentz factor                                                &$\Gamma_{\rm b}$        & 11         & 11       \\
Distance of the emission region from the black hole, in parsec & $R_{\rm dist}$ & 0.72    & 0.66     \\
Size of the BLR, in parsec                                      & $R_{\rm BLR}$              & 0.56      & 0.56      \\
Compton dominance                                            & $CD$                              & 22         & 45             \\
\hline
Jet power in electrons, in \lum, in log scale                        & $P_{\rm e}$                & 45.4            & 45.4   \\
Jet power in magnetic field, in \lum, in log scale                 & $P_{\rm B}$                & 46.2            & 46.1   \\
Radiative jet power, in \lum, in log scale                            & $P_{\rm r}$                 & 47.1            & 47.6   \\
Jet power in protons, in \lum, in log scale                          & $P_{\rm p}$                & 47.6            & 47.6   \\
\tableline
\end{tabular}
\end{center}
}
\end{table*}

\end{document}